\documentclass[prb,twocolumn,showpacs,superscriptaddress,floatfix]{revtex4-1}
\usepackage{graphicx,amsfonts,amssymb,amsmath,xspace,dsfont}
\usepackage[colorlinks=true,citecolor=green,linkcolor=red,urlcolor=blue]{hyperref}
\usepackage{color}

\newlength{\ldag}
\newcommand{\ket}[1]{|\!#1\rangle} 
\newcommand{\bra}[1]{\langle#1|}

\begin{document}

\title{Mott physics in the half-filled Hubbard model on a family of vortex-full square lattices}

\author{D. Ixert}
\affiliation{Lehrstuhl f\"ur Theoretische Physik 1, TU Dortmund, Germany}
\author{F.F. Assaad}
\affiliation{Institut f\"ur Theoretische Physik und Astrophysik, Universit\"at W\"urzburg, Am Hubland, D-97074 W\"urzburg, Germany}
\author{K.~P.~Schmidt}
\email{kai.schmidt@tu-dortmund.de}
\affiliation{Lehrstuhl f\"ur Theoretische Physik 1, TU Dortmund, Germany}

\date{\rm\today}


\begin{abstract}
We study the half-filled Hubbard model on a one-parameter family of vortex-full square lattices ranging from the isotropic case to weakly coupled Hubbard dimers. The ground-state phase diagram consists of four phases: A semi-metal and a band insulator which are connected to the weak-coupling limit, and a magnetically ordered N\'eel phase and a valence bond solid (VBS) which are linked to the strong-coupling Mott limit. The phase diagram is obtained by quantum Monte Carlo (QMC) and continuous unitary transformations (CUTs). The CUT is performed in a two-step process: Non-perturbative graph-based CUTs are used in the Mott insulating phase to integrate out charge fluctuations. The resulting effective spin model is tackled by perturbative CUTs about the isolated dimer limit yielding the breakdown of the VBS by triplon condensation. We find three scenarios when varying the interaction for a fixed anisotropy of hopping amplitudes: i) one direct phase transition from N\'eel to semi-metal, ii) two phase transitions VBS to N\'eel and N\'eel to semi-metal, or iii) a smooth crossover from VBS to the band insulator. Our results are consistent with the absence of spin-liquid phases in the whole phase diagram.
\end{abstract}


\maketitle

\section{Introduction}
%
%
The Hubbard model represents the most important microscopic model for describing solid states, since it describes on the simplest level the interplay between electronic band structure and Coulomb interactions. Despite its simplicity, the Hubbard model inhibits a huge body of physics, e.g.~it is known to contain quantum magnetism and superconducting phases. 

At half-filling, one expects typically two phases at zero temperature: a metal in the weak-coupling limit and a Mott insulator with magnetic long-range order in the domain when correlations are strong. The latter is easily understood for geometrically unfrustrated lattices, since the Hubbard model can be mapped to the corresponding Heisenberg model in the strong-coupling limit realizing a magnetically ordered ground state with broken $\text{SU}(2)$-symmetry. One promising route to more exotic Mott phases is then to introduce geometric frustration which tends to destabilize magnetic order, e.g.~one has numerical evidence for a gapped quantum spin liquid with topological order for the Heisenberg model on the highly frustrated kagom{\'e} lattice \cite{Yan10,Depenbrock12}. 

\begin{figure}
\begin{center}
\includegraphics*[width=\columnwidth]{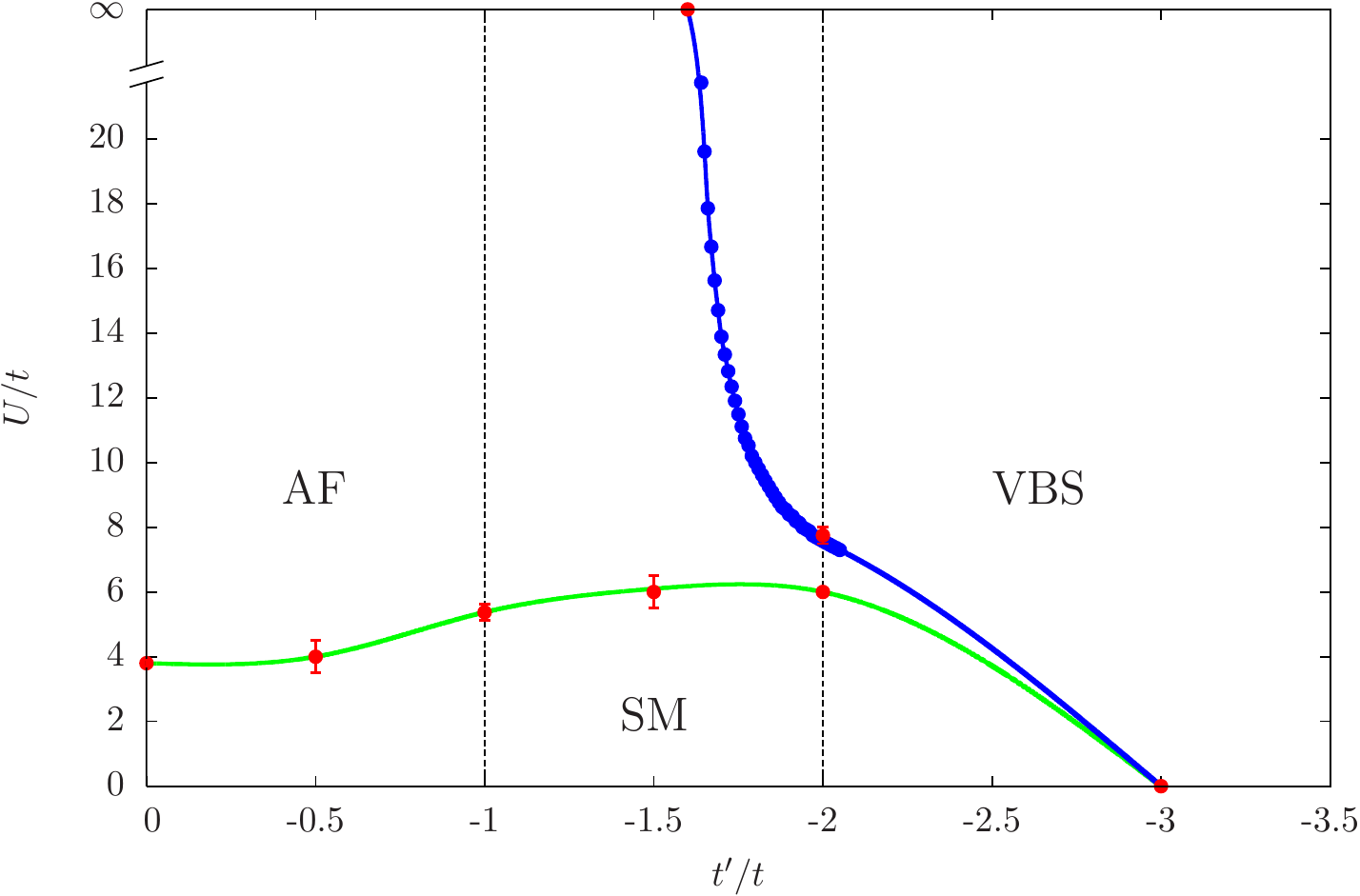}
\end{center}
\caption{(Color online) Ground-state phased diagram as a function of $t^\prime/t$ and $U/t$ as obtained from QMC (red circles) and CUT (blue circles). Solid lines are guides to the eye for the expected quantum critical lines between the N\'eel-ordered antiferromagnet (AF), the semi-metal (SM), and the valence bond solid (VBS). Dashed vertical lines denote the specific cases $t^\prime/t=-1$ and $t^\prime/t=-2$ which play an important role in this work. The QMC data at $t'/t=0$ stem from Ref.~\onlinecite{Assaad13}. The data at $t'/t=-0.5$ and $t'/t=-1.5$ result from extrapolation to infinite size of the spin-spin correlation function using lattices up to 784 sites.}  
\label{fig:phase_diagram_global}
\end{figure}

An alternative route to fascinating states of quantum matter has been pursued in recent years, by analysing two-dimensional Hubbard models in the regime of {\it intermediate} interactions. Here charge fluctuations are expected to soften the Mott insulator and there is convincing evidence that a quantum disordered phase is present on the frustrated triangular lattice \cite{Morita02,Motrunich05,Kyung06,Sahebsara08,Tocchio08,Yoshioka09,Yang10} close to the metal-insulator transition.

For each of the two scenarios to stabilize non-trivial quantum phases, geometric frustration plays an essential role. In the case of the Mott transition, geometric frustration  is crucial since it is required to protect the Fermi surface against nesting instabilities. Another possibility to avoid nesting instabilities is to drive the density of states to zero.    
This can be achieved, for example, on the $ \pi$-flux  square  or honeycomb lattices.   In this case,   it is known that the semi-metal phase is stable with respect to the Hubbard interaction.   The transition from the semi-metal to antiferromagnetic  Mott insulating state has  captured attention since numerical studies put forward the possibility of an intermediate disordered state \cite{Meng10,Chang12}.   More recent results on the honeycomb lattice \cite{Sorella12,Assaad13} exclude this possibility, and provide an explanation of the Mott transition in terms of Gross-Neveu criticality \cite{Herbut09} , where the origin of the mass gap   stems from the  symmetry breaking.    

In this work we focus on a one-parameter family of vortex-full square lattices ranging from the isotropic vortex-full square lattice just mentioned to an anisotropic limit of weakly coupled Hubbard dimers. We combine CUTs and QMC simulations to determine the ground-state phase diagram which we display in Fig.~\ref{fig:phase_diagram_global}. The latter consists entirely of a semi-metal phase, a band insulator, a magnetically ordered N\'eel state, and a VBS. Note that BI and VBS are in fact continuously connected and therefore not distinct phases. It is nevertheless useful to introduce both terms when considering the weak- or strong-coupling limit. We establish i) a direct transition from N\'eel order to semi-metal for the isotropic vortex-full square lattice, ii) a crossover from VBS to band insulator if dimerization is large enough, and iii) the existence of two quantum phase transitions from VBS to N\'eel and N\'eel to semi-metal in the intermediate regime. Altogether, our results are consistent with the absence of spin-liquid phases in the whole phase diagram.

We organize the paper as follows. In Sec.~\ref{sec:model} we define the Hubbard model and the lattice under consideration. Additionally, we consider the extreme limits of free fermions and of strong coupling yielding the presence of at least four phases in the phase diagram. Afterwards, we present our two-level CUT approach in Sec.~\ref{sec:cut} and we describe technical aspects of the QMC simulations in Sec.~\ref{sec:qmc}. The full body of results including the ground-state phase diagram is given in Sec.~\ref{sec:results}. Finally, we summarize our results in Sec.~\ref{sec:conclusion}.
%
%
\section{Model}
\label{sec:model}
%
%
We study the Hubbard model
\begin{eqnarray}
\mathcal{H} &=& \mathcal{H}_U+\mathcal{H}_t\nonumber\\
            &=& U\sum_{i}n_{i\uparrow}n_{i\downarrow} -\sum_{\langle i,j\rangle,\sigma}t_{ij}\left(c_{i\sigma}^{\dagger}c ^{\phantom{\dagger}}_{j\sigma}+\textrm{h.c.}\right)
\label{Hubbard_model}
\end{eqnarray}
at half-filling and zero temperature. The occupation number operator for fermions with spin $\sigma$ at the site $i$ of the lattice is $n_{i\sigma}=c_{i\sigma}^{\dagger}c^{\phantom{\dagger}}_{i\sigma}$. The hopping amplitudes $t_{ij}$ are chosen such that each plaquette contains a $\pi$-flux. Here we consider in each plaquette three identical hopping amplitudes $t>0$ and one coupling $t^\prime<0$ arranged as illustrated in Fig.~\ref{fig:lattice}, if not stated otherwise. In the following we denote by $b_t$ ($b_{t^\prime}$) bonds with an hopping amplitude $t$ ($t^\prime$). 

Our aim is to determine the ground-state phase diagram as a function of $t/U$ and $t^\prime/t\leq -1$. The case $t^\prime/t=-1$ corresponds to the isotropic vortex-full square lattice while $t^\prime/t\rightarrow -\infty$ represents the limit of isolated Hubbard dimers on $t^\prime$ bonds which build a staggered pattern. 

\subsection{Non-interacting case}
\label{ssec:free}

The limit $U=0$ of free fermions can be solved exactly. Our lattice has two sites per unit cell which we take here as the two sites of bonds $b_{t^\prime}$.  
Applying a Fourier transformation   yields the Hamiltonian
\begin{equation}
\mathcal{H_{\rm free}} = \sum_{\vec{k},\sigma}     \left( \tilde{a}_{\vec{k},\sigma}^{\dagger}, \tilde{b}_{\vec{k},\sigma}^{\dagger}   \right) 
\underbrace{
\begin{pmatrix}
0   & Z_{\vec{k}} \\
Z_{\vec{k}}^{\dagger} & 0
\end{pmatrix}}_{H(\vec{k})}
\begin{pmatrix}
\tilde{a}_{\vec{k},\sigma} \\ 
\tilde{b}_{\vec{k},\sigma}  
\end{pmatrix}\,.
\label{free_model}
\end{equation}
The two energy bands then read
\begin{eqnarray}
E_{\pm}(\vec{k})  &=&  \pm  \left\| Z_{k} \right\|\\   
                 &=&   \pm  \left\| \left(  t^\prime - 1\right)  +   \left( 1 +  {\rm e}^{- {\rm i}\vec{k} \cdot \vec{a}_2} \right)  \left(  1 +  {\rm e}^{-{\rm i} \vec{k} \cdot \vec{a}_1} \right)  \right\|\nonumber
\label{free_model_bands}
\end{eqnarray}
with the lattice vectors $\vec{a}_1 = a(1,1)$ and $\vec{a}_2 = a(1,-1)$, and setting $t=1$.  At the particle-hole symmetric point and in the regime 
$-3< t^\prime< 1$,  the  Fermi surface consists of two points. Using $\vec{b}_{i}\cdot \vec{a}_{j}=\delta_{i,j}$, one finds 
\begin{equation}
	\vec{K}_{\pm}   =  \pm \cos^{-1}\left(   \frac{ -t^\prime -1 }{2} \right)  \left(  \vec{b}_{1} - \vec{b}_{2} \right)\,.  
\end{equation}
Linearization around the two Fermi points  gives two two-component Dirac cones,
\begin{equation}
\begin{gathered}
H(\; \vec{K}  +  \vec{p}  )  =    - v_x p_x \sigma_y  +  v_y p_y \sigma_x    \\ 
H( - \vec{K}  +  \vec{p}  )  =    \;   v_x p_x \sigma_y  -  v_y p_y \sigma_x    
\end{gathered}
\end{equation}
 with opposite vorticities and with velocities $ v_x = t^\prime -1 $, $v_y  = - \sqrt{ 4 - \left(t^\prime + 1 \right)^2} $.  Here 
 $ \sigma_x,\sigma_y $ correspond to the Pauli spin matrices. 
In the range  $  -3 < t^\prime   < 1 $,   the velocities are finite. In this regime, in  the continuum limit and at the single-particle 
level, one can show that  this  state of matter can  acquire  a mass gap only  by breaking symmetries \cite{Ryu09}. 

We now comment on special values of $t^\prime$:  At $t^\prime=1$,  both velocities vanish and the Fermi surface is that of the square lattice with van-Hove singularity. At $t^\prime=0$ and $t^\prime=-1$, one recovers the honeycomb and $\pi$-flux square lattices respectively. Both values of $t^\prime$ lead to enhanced lattice symmetry (C$_3$ at $t^\prime=0$ and C$_4$ at \mbox{$t^\prime=-1$}) which has the effect of pinning the location of the Dirac points.  At $t^\prime = -3$,  the two Dirac points merge at the $\Gamma$-point and one  of the Fermi velocities vanishes  thereby yielding a single-particle density of states $N(\omega) \propto \sqrt{\omega} $. Beyond $t^\prime = -3$, a single-particle gap opens and the state is adiabatically connected to independent dimers. 
 
Another quantity of interest is the bandwidth $W$ which is a characteristic scale for the kinetic energy of the electrons. One finds
\begin{eqnarray}
W &\equiv & \max_{\vec{k}}\left(E_{+}({\vec{k}})\right) - \min_{\vec{k}} \left( E_{-}({\vec{k})} \right)\nonumber\\
 &=& 2\sqrt{(t^\prime-t)^2+4t^2}\, ,
\label{free_model_band_width}
\end{eqnarray}
which gives $4\sqrt{2}t$, $2\sqrt{13}t$, and $4\sqrt{5}t$ for \mbox{$t^\prime/t\in\{-1,-2,-3\}$}.

\subsection{Strong-coupling limit}
\label{ssec:strong_coupling}
In the strong-coupling limit $t,t^\prime\ll U$, charge excitations are frozen and the low-energy physics of the Hubbard model is determined by spin degrees of freedom. An effective low-energy spin model can be derived by degenerate perturbation theory, and one obtains in leading order a $J$-$J^{\prime}$-Heisenberg model
\begin{equation}
 \mathcal{H}_{\rm spin}=E_0+J\sum_{b_t}\vec{S}_i\cdot\vec{S}_j + J^{\prime}\sum_{b_{t^\prime}}\vec{S}_i\cdot \vec{S}_j
\label{Heisenberg_model}
\end{equation}
with energy per site $E_0/N\,$=$\,-(3t^2+{t^\prime}^2)/(2U)$, \mbox{$J=4t^2/U$}, and \mbox{$J^{\prime}=4{t^\prime}^2/U$}. 

This spin model is unfrustrated and can therefore be simulated efficiently by QMC \cite{Wenzel08}. One finds a second-order phase transition at \mbox{$J^{\prime}/J\approx 2.5196$} which corresponds to \mbox{$t^\prime/t=\pm\sqrt{J^{\prime}/J}\approx \pm 1.5873$}. This critical point separates a magnetically ordered N\'eel phase with broken $\text{SU}(2)$-symmetry and gapless Goldstone bosons at smaller ratios $J^{\prime}/J$ from a paramagnetic VBS with gapped triplon excitations. The universality class of this transition is known to be O(3)\cite{Wenzel08}.
  
\begin{figure}[ht]
\includegraphics*[width=0.7\columnwidth]{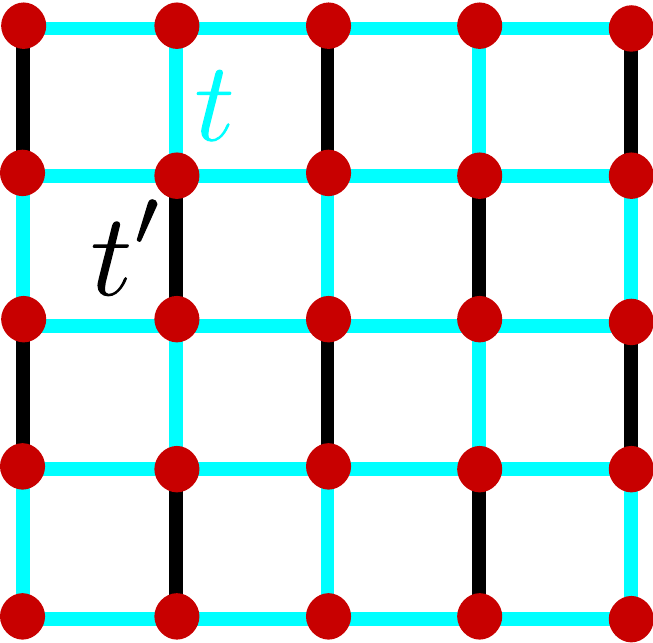}
\caption{(Color online) Illustration of the $t$-$t^\prime$ Hubbard model. Filled circles denote sites and dark (gray) lines refer to the coupling $t^\prime$ ($t$).} 
\label{fig:lattice}
\end{figure}
%
%
\section{CUTs}
\label{sec:cut}
%
In this section, we present our two-step CUT approach. First, we use graph-based CUTs (gCUTs) to derive an effective low-energy spin model in the Mott phase of the Hubbard model by separating spin and charge degrees of freedom along the lines of Refs.~\onlinecite{Yang11,Yang12}. Afterwards, we apply perturbative CUTs (pCUTs) to derive high-order series expansions for the VBS phase of the effective spin model.
%
\subsection{gCUT}
\label{ssec:gcut}
%
The goal of CUTs~\cite{Wegner94,Glazek93,Glazek94} is to transform the Hamiltonian into an optimized basis representation. For the Hubbard model in the Mott insulating phase, this basis brings the Hamiltonian in a blockdiagonal form where charge and spin degrees of freedom are decoupled.

To this end the Hamiltonian is considered as a continuous function $\mathcal{H}(l)$ of the flow parameter $l$ with \mbox{$\mathcal{H}(l=0)=\mathcal{H}$} and \mbox{$\mathcal{H}(l=\infty)=\mathcal{H}_\text{eff}$} as the starting Hamiltonian and the effective Hamiltonian respectively. With an antihermitian generator $\eta(l)$ one gets the flow equation
\begin{equation}
\frac{\text{d}\mathcal{H}(l)}{\text{d}l}=\left[\eta(l), \mathcal{H}(l)\right] \; . \label{eq:flowEquation}
\end{equation}

Here we separate states without double occupancies ($0$DO states) from the states with one or more double occupancies ($n$DO states with $n>0$), so we choose the quasi-particle generator~\cite{Knetter00,Mielke98,Knetter03}
\begin{equation}
\eta_{i,j}(l)=\text{sgn}(q_i-q_j) \mathcal{H}_{i,j}(l)
\end{equation}
in an eigenbasis of a counting operator $Q$. The eigenvalues of $Q$ are $q_i=0$ for $0$DO states and $q_i=1$ for states with DOs. As $[Q,\mathcal{H}_{\rm eff}]=0$, the effective Hamiltonian consists of a decoupled block without DOs, i.e.~spin and charge degrees of freedom have been separated. This allows us to derive an effective spin model as we will detail below. We stress that our choice for $Q$ is different from the usual choice which identifies $Q$ with the number operator of DOs \cite{MacDonald88,Reischl04,Hamerla10,Yang11,Yang12}. Although in both cases one is able to derive an effective spin model, our definition of $Q$ has numerical advantages, since the corresponding CUT has to uncouple less operators.     

The commutator in the flow equation \eqref{eq:flowEquation} leads typically to an infinite number of terms, so that a truncation must be performed. The truncation scheme used in gCUTs \cite{Yang11} is to solve the flow equation on topologically distinct finite clusters, called graphs. On these graphs the Hamiltonian has a representation in terms of a finite matrix. Consequently, the flow equation has to converge and can be solved numerically exact. Afterwards, one subtracts all subgraph contributions from a given graph. These reduced contributions of each graph are embedded on the infinite lattice to get a result in the thermodynamic limit. The number of possible embeddings is called embedding factor.

\begin{figure}
\begin{center}
\includegraphics*[width=0.95\columnwidth]{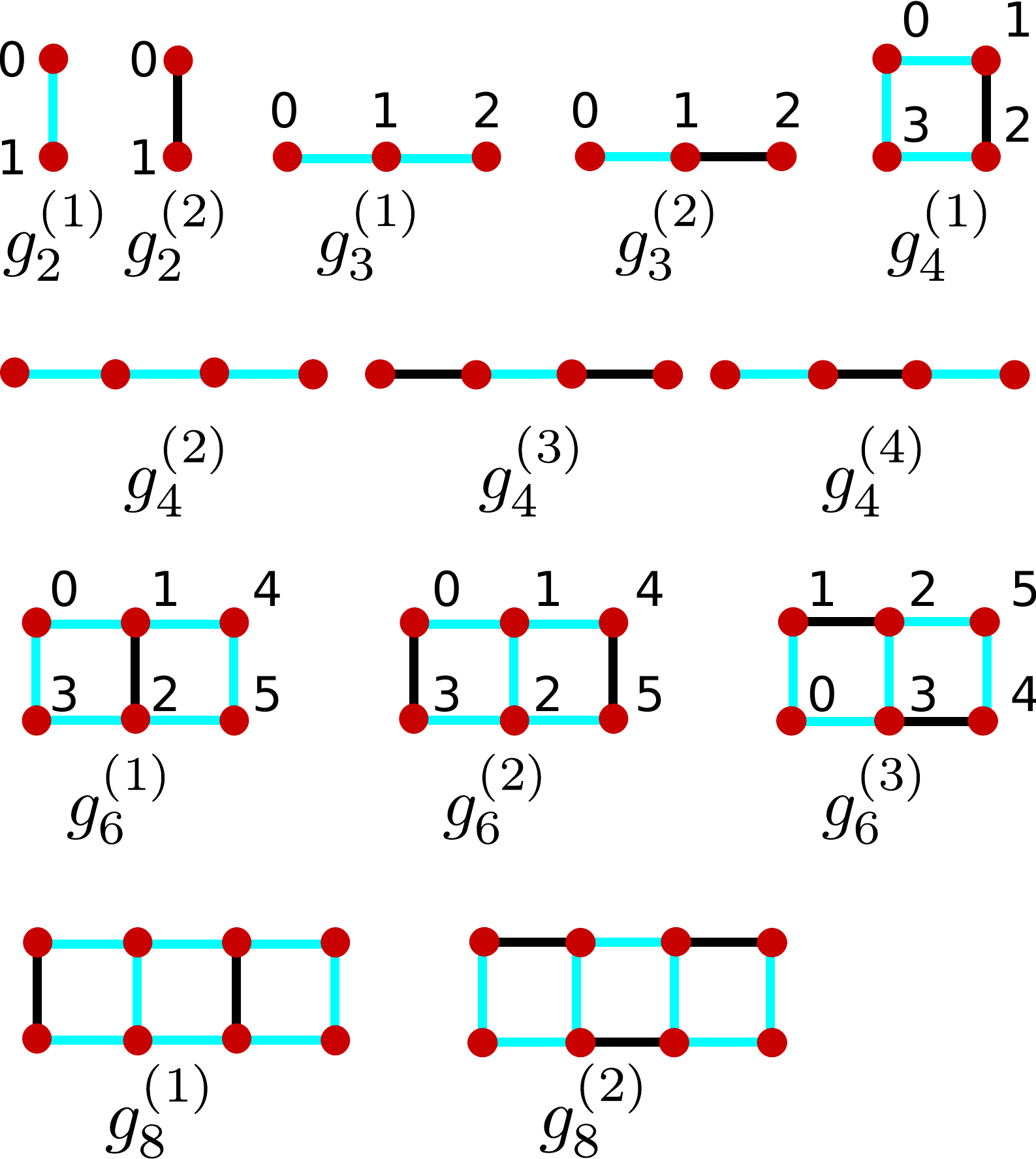}
\end{center}
\caption{(Color online) Illustration of graphs $g_\nu^{(n)}$ used in the gCUT plaquette expansion. The index $\nu$ classifies the graphs by the number of Hubbard sites which are then further counted by $n=1,2,3,\ldots$\,\,. Red filled circles denote Hubbard sites which are connected via hopping amplitudes $t^\prime$ ($t$) by black (cyan) lines.} 
\label{fig:allGraphs}
\end{figure}

The numerical effort of the gCUT depends essentially on the graph size as well as on the total number of graphs, since both grow exponentially with the number of sites. A full graph decomposition of the lattice meets the additional challenges that single graphs can contain less lattice symmetries compared to the full problem and the embedding factors might become very large demanding a very high numerical precision. Therefore, we decided not to perform a full graph decomposition, but to expand in terms of rectangular graphs \cite{Dusuel10} which keeps the number of graphs and the embedding factors smaller (see Ref.~\onlinecite{Yang12} for a similar approach on the honeycomb lattice). Here we calculated all contributions from the topological distinct graphs $g_\nu^{(n)}$ up to three plaquettes shown in Fig.~\ref{fig:allGraphs} having up to $\nu=8$ Hubbard sites. Furthermore we neglected all graphs which have more than $4$ sites in one direction. The index $n$ is used to numerate clusters with the same number of sites $\nu$.
Furthermore we define the gCUT calculation with a maximal cluster of $\nu$ sites as $\text{gCUT}(\nu)$.
%
%
\subsubsection{Symmetries}
\label{sssec:gcut_symmetries}
In principle one has to create all possible Hubbard states on a particular graph and solve the graph-dependent CUT. But in order to efficiently solve the flow equations one should keep the size of the matrix as small as possible, since the total number of flow equations of $g_\nu^{(n)}$ increases quickly as $4^\nu\times4^\nu$. This can be achieved by implementing symmetries. In Refs.~\onlinecite{Yang11,Yang12} the only symmetry taken into account is the conservation of $S^z_{\text{total}}$. Here we go beyond this scheme:

First, as acting with the Hamiltonoperator on states conserves all symmetries, it is useful  i) to choose the 0DO states as initial states and ii) to act with $\mathcal{H}$ on these states to create other orthogonal states containing DOs using the Gram-Schmidt algorithm. After each acting with $\mathcal{H}$, one performs the CUT and checks if the 0DO block is already converged after the CUT. In contrast to creating all states of a given graph, we observe this approach to converge with already much less states compared to the full Hilbert space dimension.

Second, we use the {\it full} $\text{SU}(2)$-symmetry of the Hubbard model when constructing the 0DO states.
This can be achieved by creating all half-filled Hubbard states without DOs as spin states. 
Using the $\text{SU}(2)$-Symmetry is then only a matter of adding angular momenta through 
generalized Clebsch-Gordan-coefficients~\cite{Schnalle2010}. 

Therefore, the CUT on graph $g_\nu^{(n)}$ is performed for each pair of quantum numbers $(S,S^z)$ independently
 with an optimal basis  which allows us to reduce the numerical effort considerably. 

For the CUT itself the convergence criteria is defined with respect to the so called residual off-diagonality (ROD), which is the sum of squared off-diagonal elements. All elements which couple the 0DO to the $n$DO-block are squared and summed over. As soon as the ROD reaches a value lower than $10^{-10}$ the CUT is stopped.
The quasi-particle generator leads typically to a minimization of the ROD, but it can be a problem if the $n$DO-block contains eigenvalues which lie below the eigenvalues of the 0DO-block. In this situation the quasi-particle generator exchanges the corresponding levels which implies that a separation of spin and charge degrees of freedom is not straightforward on the corresponding graph. It is then reasonable to abort the CUT if the ROD has reached a minimal value, since the exchange of levels during the flow is typically accompanied with a temporarilly increasing ROD. This strategy breaks down if the remaining ROD is still sizable which occurs for the current problem at large values of $t/U$ as detailed below. 

The basis is generated by successively acting with the Hamiltonian on all 0DO states. This is performed until the difference in the 0DO-block is below $10^{-8}$.
For example, on $g^{(1)}_6$ at $t^\prime/t=-2$ and $t/U=0.1$, the number of spin states and the number of required states for convergence of the 0DO-block is shown in Tab.~\ref{tab:spinStates}. Clearly, one needs much less states for convergence if the full $\text{SU}(2)$-symmetry is taken into account instead of only using the $S^z$-symmetry.
\begin{table}
\begin{tabular}{ccc||ccc}
$S$ & 0DO & required & $S^z$ & 0DO & required \\\hline
 $S=0$ & $5$ & $48$ & $S^z=0$ & $20$ & $400$ \\
 $S=1$ & $9$ & $73$ & $S^z=\pm1$ & $15$ & $225$ \\
 $S=2$ & $5$ & $19$ & $S^z=\pm2$ & $6$ & $36$ \\
 $S=3$ & $1$ & $1$ & $S^z=\pm3$ & $1$ & $1$ \\
\end{tabular}
\caption{Number of spin-states (0D0 states) and required states for convergence of the 0DO-block by using $\text{SU}(2)$-Symmetry or $S^z$-Symmetry on $g_6^{(1)}$ for $t^\prime/t=-2$ and $t/U=0.1$.}
\label{tab:spinStates}
\end{table}

\subsubsection{Effective Spin model}
\label{sssec:effectiveSpinModel}
After the CUT has been done for all graphs $g\equiv g_\nu^{(n)}$, one obtains for each $g$ 
an effective Hamiltonmatrix $\mathcal{H}^g_\text{eff}$ containing two decoupled blocks. The block
 without DOs is used to derive an effective graph-dependent spin model in an operator basis
\begin{eqnarray}
 \mathcal{H}_\text{spin}^{g} = E^g_0 &+& \sum_{i,j} J_{ij}^g \,\,\,\vec{S}_i\cdot \vec{S}_j \nonumber\\
  &+& \sum_{i,j,k,l} J^g_{ijkl}\,\left(\vec{S}_i \cdot\vec{S}_j\right) \left( \vec{S}_k \cdot\vec{S}_l\right)+ \ldots
\end{eqnarray}
where $E^g_0$ denotes a constant, $J_{ij}^g$ Heisenberg couplings, and $J^g_{ijkl}$ four-spin interactions. The $"\ldots"$ refer to $n$-spin operators with $n\in\{6,8,\ldots\}$. The spin couplings are calculated by demanding
\begin{equation}
\bra{i}\mathcal{H}^g_\text{eff}\ket{j} = \bra{i}\mathcal{H}^g_\text{spin}\ket{j} \; ,
\end{equation}
which leads to an overdetermined equation system for the spin couplings.

From the effective spin models on each graph an effective spin model in the thermodynamic limit can be derived by embedding the reduced spin couplings in the infinite lattice. The reduced spin couplings $\overline{J^g}$ of graph $g$ are obtained by subtracting all contributions from subgraphs $f\subset g$ 
\begin{align}
 \overline{J^g_{\phantom{-}}} = J^g - \sum_{f\subset g} \overline{J^f} \; .
\end{align}
As an example, let us consider the important nearest-neighbor Heisenberg coupling $J^\prime$ on $b_{t^\prime}$-bonds (see Fig.~\ref{fig:lattice}). The reduced contributions up to $\text{gCUT}(6)$ are determined by
\begin{align}
 \overline{J_{01}^{g_2^{(2)}}} &= J_{01}^{g_2^{(2)}} \nonumber\\
 \overline{J_{12}^{g_3^{(2)}}} &= J_{12}^{g_3^{(2)}} - \overline{J_{01}^{g_2^{(2)}}}\nonumber \\
 \overline{J_{12}^{g_4^{(1)}}} &= J_{12}^{g_4^{(1)}} - \overline{J_{01}^{g_2^{(2)}}} \nonumber\\
 \overline{J_{12}^{g_6^{(1)}}} &= J_{12}^{g_6^{(1)}} - \overline{J_{01}^{g_2^{(2)}}} - 2\overline{J_{12}^{g_4^{(1)}}}\nonumber  \\
 \overline{J_{03}^{g_6^{(2)}}} &= J_{03}^{g_6^{(2)}} - \overline{J_{01}^{g_2^{(2)}}} - \overline{J_{12}^{g_4^{(1)}}}  \nonumber\\
 \overline{J_{12}^{g_6^{(3)}}} &= J_{12}^{g_6^{(3)}} - \overline{J_{01}^{g_2^{(2)}}} - \overline{J_{12}^{g_4^{(1)}}} - \overline{J_{12}^{g_3^{(2)}}}  
\end{align}
for the graphs displayed in Fig.~\ref{fig:allGraphs}. The overall exchange $J^{\prime}$ is then given by summing over these reduced contributions weighted by the appropriate embedding factors
\begin{equation}
 J^\prime = \overline{J_{01}^{g_2^{(2)}}} + 2\overline{J_{12}^{g_3^{(2)}}} + 2\overline{J_{12}^{g_4^{(1)}}} +\overline{J_{12}^{g_6^{(1)}}} +2\overline{J_{03}^{g_6^{(2)}}} +4\overline{J_{12}^{g_6^{(3)}}} \; .
\end{equation}
The same kind of procedure has to be performed for all spin operators which fit on the considered graphs. Here we stop with graphs containing up to three plaquettes, i.e.~one has two-spin, four-spin, six-spin, and eight-spin interactions in the effective spin Hamiltonian in the thermodynamic limit.

\begin{figure}[ht]
\includegraphics*[width=0.85\columnwidth]{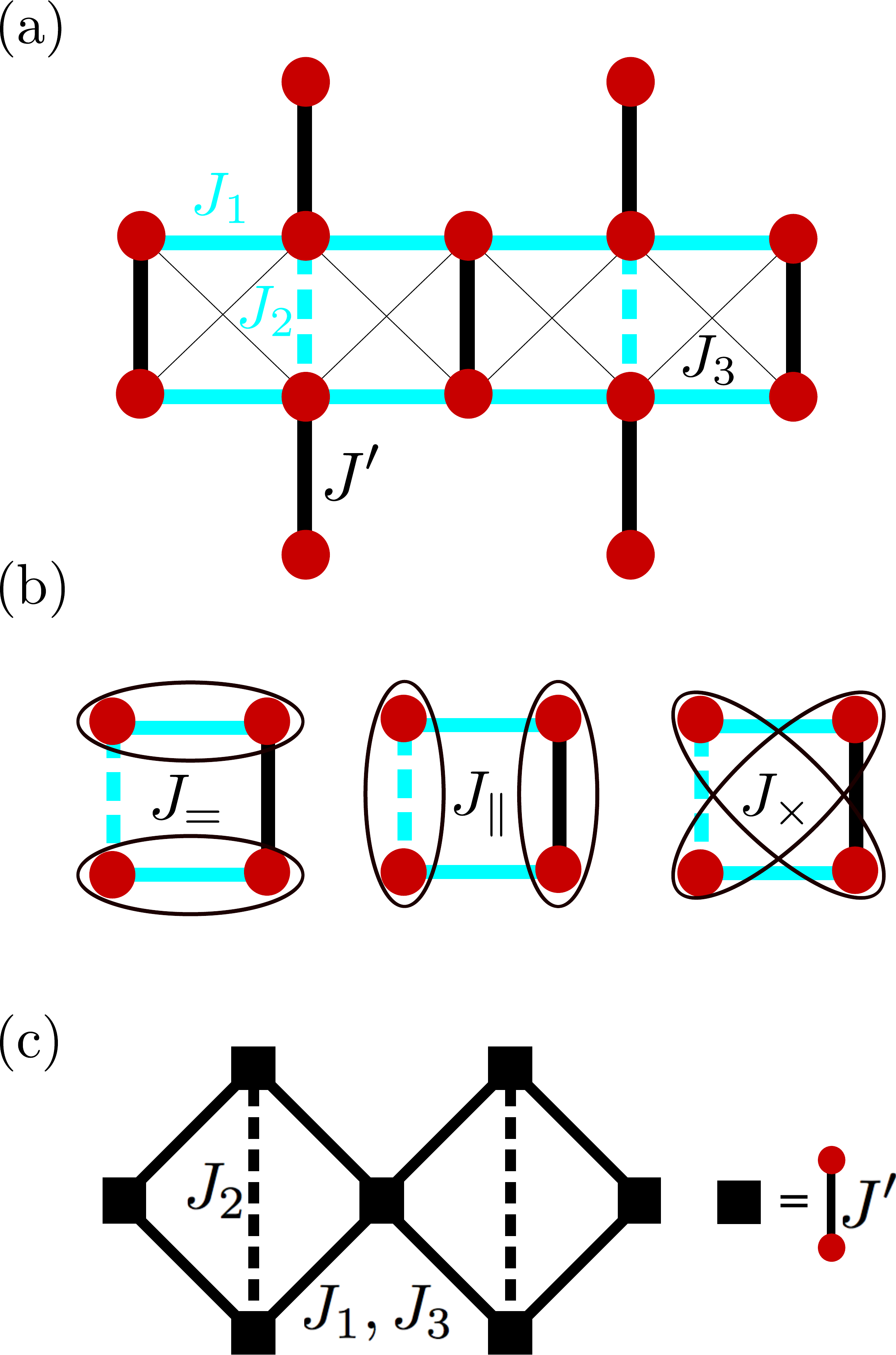}
\caption{Illustration of (a) two-spin and (b) four-spin interactions in the original lattice. (c) The effective lattice resulting from replacing $t^\prime$-bonds by effective supersites (filled black squares) together with the corresponding two-spin interactions as links.} 
\label{fig:eff_lattice}
\end{figure}

In this work we restrict the discussion to all two-spin and four-spin interactions living on single plaquettes as illustrated in Fig.~\ref{fig:eff_lattice}(a)-(b): First, the size of the exchange couplings depends on $t/U$ and $t^\prime/U$. Therefore, we include almost all couplings arising up to order four pertubation theory in the strong-coupling limit. The only exception is a next-nearest neighbor Heisenberg interaction which is known to have a small amplitude. Second, we include four-spin ring exchange interactions on plaquettes which are very large and important for the Hubbard model on the isotropic square and triangular lattice \cite{MacDonald88,Reischl04,Hamerla10,Yang11,Yang12}.
The coupling $J_3$ is one order lower in magnitude than the other included two-spin interactions. Six-spin terms are not included as they are also at least one order smaller in magnitude than the considered couplings (except for $J_3$) in most of the parameter regime considered.

The corresponding spin Hamiltonian reduces to the form
\begin{equation}
 \mathcal{H}_\text{spin} = E_0 + \mathcal{H}^\prime+ \mathcal{H}\label{eq:spinModelForm}
\end{equation}
with
\begin{align}
\mathcal{H}^\prime &= J^\prime \sum_{\langle i,j\rangle^\prime} S_i S_j \\
\mathcal{H} &= J_1 \sum_{\langle i,j\rangle_1} S_i S_j +J_2 \sum_{\langle i,j\rangle_2} S_i S_j + J_3 \sum_{\langle\langle i,j\rangle\rangle} S_i S_j\\
 &+ \sum_{\left\{i,j,k,l\right\}\in\text{Plaq.}}\Big( J_\parallel \left(S_i S_l\right)\left(S_j S_k\right) \\
  &\qquad\qquad\qquad+ J_= \left(S_i S_j\right)\left(S_k S_l\right)\\ &\qquad\qquad\qquad+ J_\times \left(S_i S_k\right)\left(S_j S_l\right)\Big) \; .
\end{align}

%
%

\begin{figure}
\begin{center}
\includegraphics*[width=0.9\columnwidth]{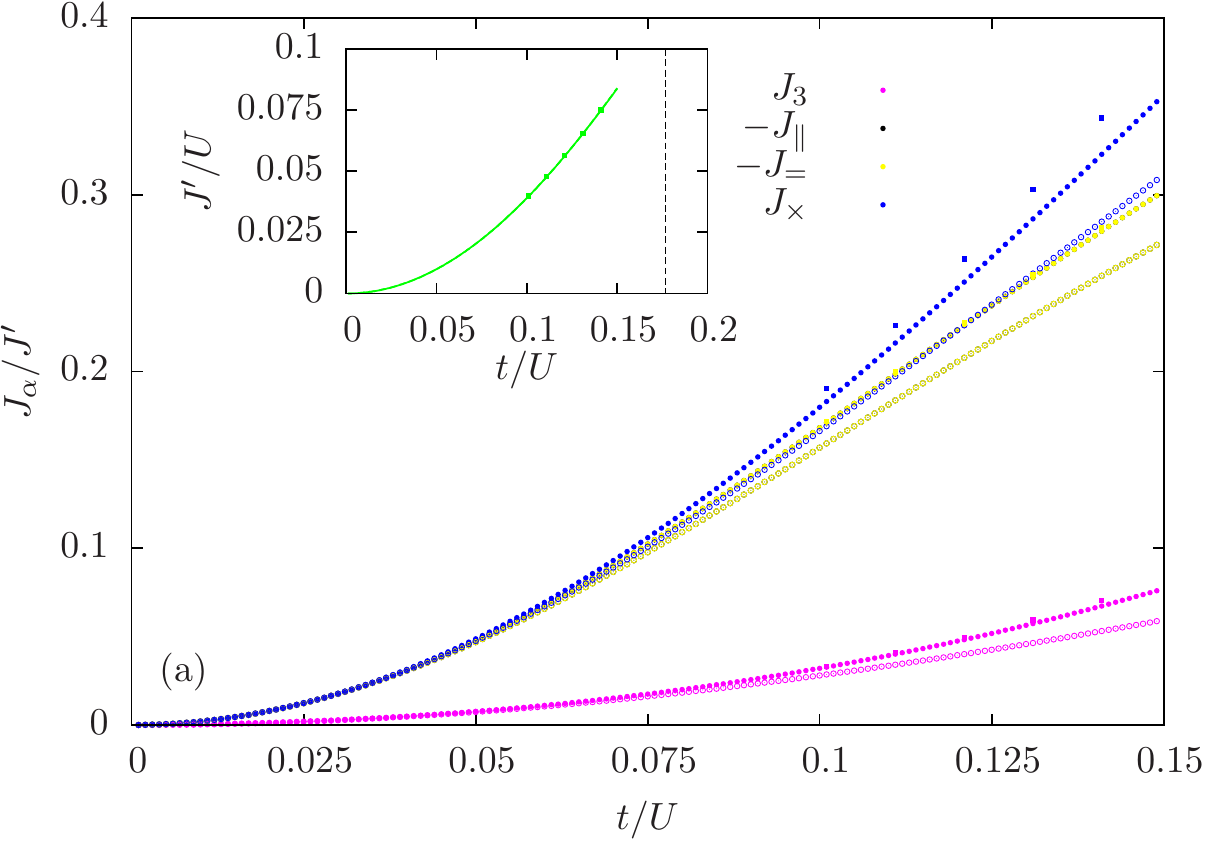}
\includegraphics*[width=0.9\columnwidth]{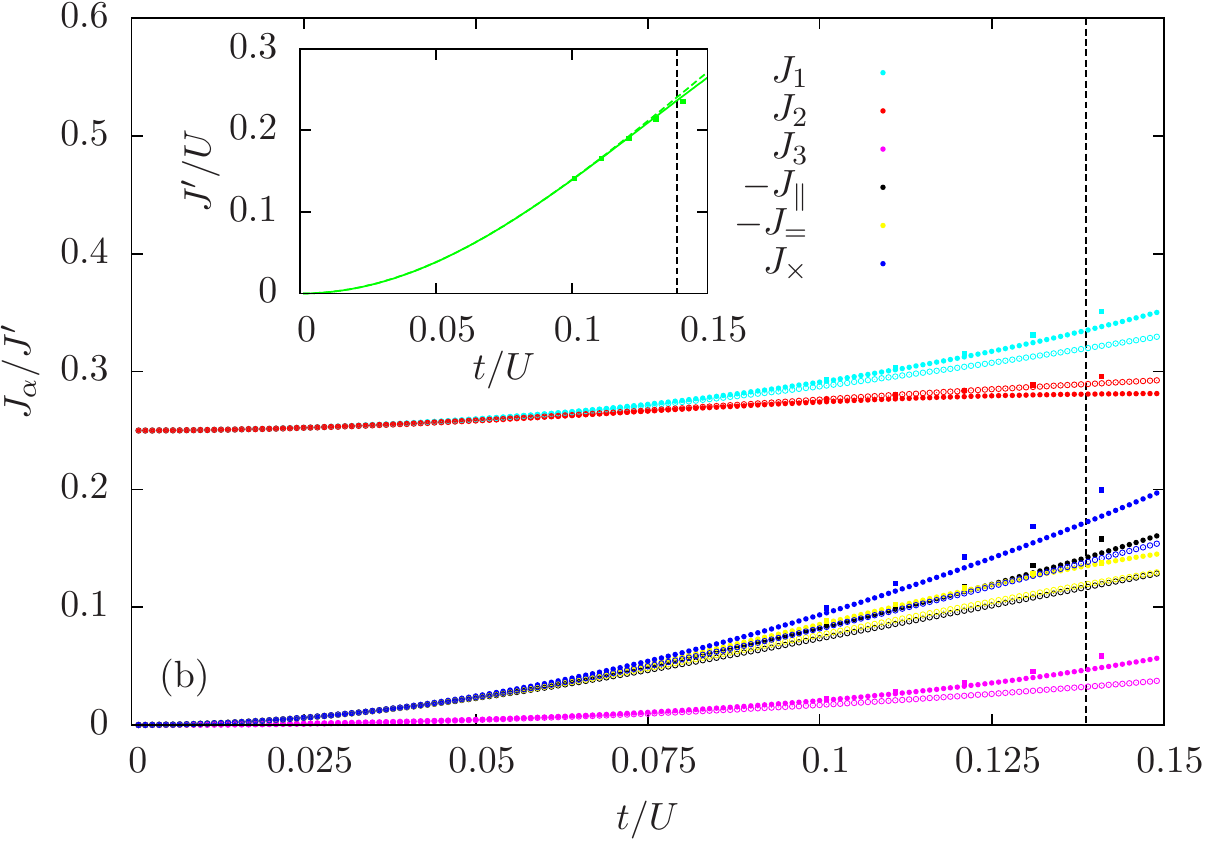}
\includegraphics*[width=0.9\columnwidth]{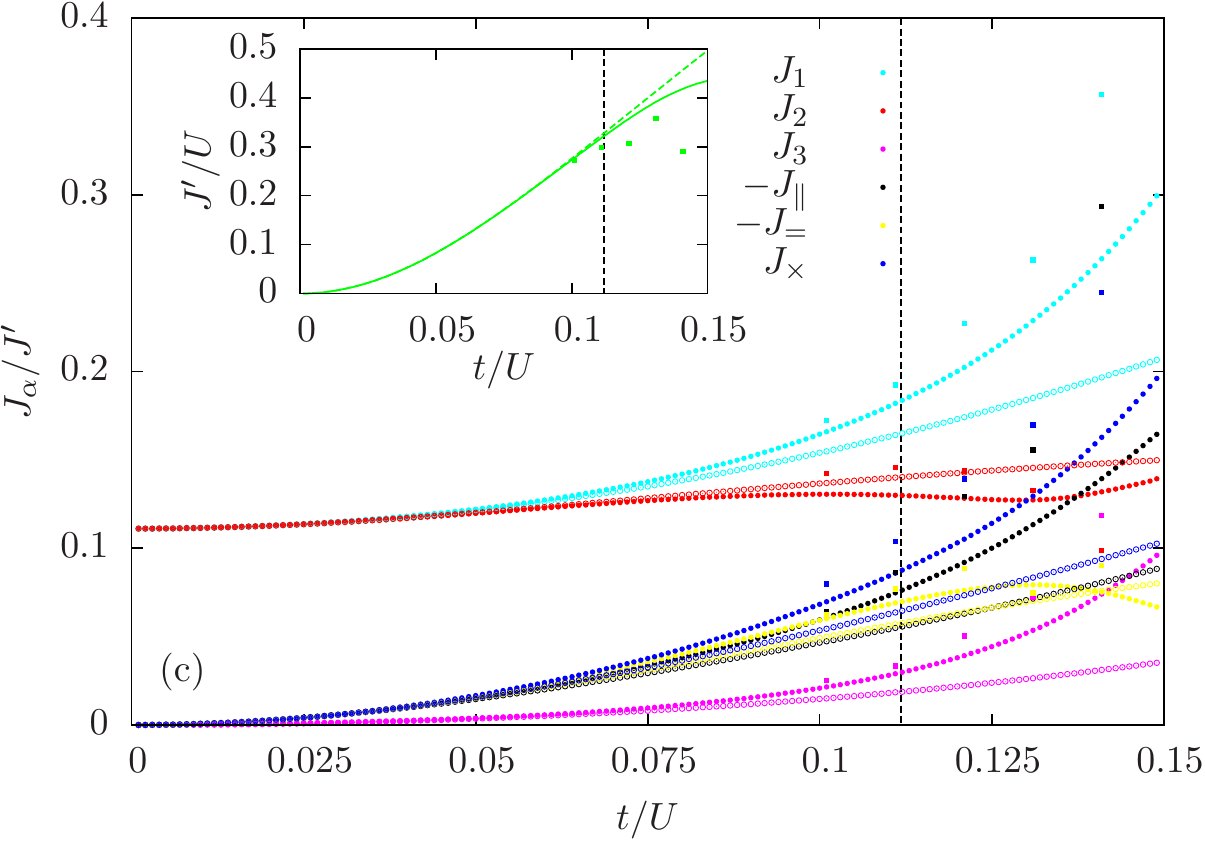}
\end{center}
\caption{(Color online) Relative values for the effective spin-couplings $J_\alpha/J^{\prime}$ with $\alpha\in\{1,2,3,\parallel,=,\times\}$ as a function of $t/U$ in the thermodynamic limit for (a) $t^\prime/t=-1$, (b) $t^\prime/t=-2$, and (c) $t^\prime/t=-3$. Empty (filled) circles are $\text{gCUT}(4)$ ($\text{gCUT}(6)$), filled squares are $\text{gCUT}(8)$. Vertical dashed lines signal the inverse bandwidth $t/W$ according to Eq.~\ref{free_model_band_width}. {\it Insets}: Exchange coupling $J^{\prime}/U$ as a function of $t/U$. Note that in (a) $J^{\prime}=J_1=J_2$.} 
\label{fig:effCouplings}
\end{figure}

The resulting strength of the magnetic exchange couplings is displayed for $t^\prime/t\in\{-1,-2,-3\}$ in Fig.~\ref{fig:effCouplings} using gCUT($\nu$) for $\nu\in\{4,6,8\}$. In all cases we observe that the inverse bandwidth $t/W$ sets a characteristic energy scale with regards to the convergence of the gCUT. For $t/U\leq t/W$ we find a continuously improving behaviour when increasing the truncation order $\nu$. In contrast, for larger values of $t/U$, we observe that even larger values of $\nu$ are required which is reasonable due to the increasing correlation length of charge fluctuations. Additionally, for the large clusters with $\nu=8$ and for the large value of $t^\prime/t=-3$, we have to abort the flow of the gCUT in the minimum of the ROD before the separation of spin and charge degrees of freedom is completed. As a consequence, the effective exchange couplings display a non-monotonic behaviour suggesting the breakdown of the applied gCUT approach in this regime.
%
%

Finally, we discuss the relative strength of the displayed couplings in the regime $t/U\leq t/W$. Here we observe the following hierarchy: nearest-neighbor two-spin interactions $J^{\prime}$, $J_1$, and $J_2$ -- arising in order two perturbation theory -- represent the dominant couplings. The most important subleading terms  are the four-spin interactions $J_\parallel$, $J_=$, and $J_\times$ located on single plaquettes. Therefore, our results are in full agreement with similar calculations for the isotropic Hubbard model on the triangular \cite{Yang11,Yang12} and the square lattice \cite{MacDonald88,Reischl04,Hamerla10}.

%
\subsection{pCUT in the VBS}
\label{ssec:pcut}
%
In the previous section we used gCUTs to derive the effective low-energy spin model Eq.~\eqref{eq:spinModelForm}. This spin model is expected to contain the low-energy physics of the original Hubbard model as long as $t/U\leq t/W$, i.e.~in the regime where charge fluctuations are not too strong. Nevertheless, the solution of the derived quantum spin model is still a very hard problem. In the strong-coupling limit $t/U\rightarrow 0$, one expects a long-range ordered N\'eel state for $t^\prime/t\leq -1.5873$ and a VBS for larger values of $t^\prime/t$ as detailed in Sect.~\ref{ssec:strong_coupling}. One important question we address in this work is if there is a quantum phase transition inside the Mott insulator between VBS and N\'eel order as a function of $t/U$ for fixed $t^\prime/t$. This can be done by studying the breakdown of the gapped VBS, i.e.~one expects the one-triplon gap to close at the quantum critical point. 

The ground state of the VBS is adiabatically connected to the limit of isolated dimers on $b_{t^\prime}$ bonds. It is therefore possible to set up a high-order series expansion about this dimer limit. Consequently, we introduce the expansion parameters $x_\kappa=J_\kappa/J^{\prime}$ with $\kappa\in\{1,2,3,\times,=,\parallel\}$ and we express the effective spin model as
\begin{equation}
 \frac{\mathcal{H}_\text{spin}}{J^{\prime}} = \frac{1}{J^{\prime}}\left( E_0 + \mathcal{H}^\prime\right)+ \sum_\kappa x_\kappa\mathcal{H}^{(\kappa)}\quad,\label{eq:spinModelForm_alpha}
\end{equation}
such that $x_\kappa=0$ corresponds to the limit of isolated dimers.

In the limit $x_\kappa=0$ the product state of singlets on all $b_{t^\prime}$-bonds becomes the exact ground state. Elementary excitations are local triplets with total spin one and excitation energy $J^{\prime}$. Operators proportional to $x_\kappa$ give rise to hoppings, interactions, or particle creation and annihilations of triplets. In the following we apply a pCUT to enforce a quasi-particle description in terms of triplons which are dressed triplet excitations and represent the elementary excitations of gapped VBS phases \cite{Schmidt03}. Replacing dimers of the original lattice by an effective site $\nu$, one finds an effective triangular lattice as shown in Fig.~\ref{fig:eff_lattice}(c).

The term $\mathcal{H}'$ is therefore the unperturbed part in the pCUT calculation. It acts locally on the effective triangular lattice and it is diagonal in the triplet-counting operator defined as
\begin{equation}
 Q_{\rm Triplon}=\sum_{\nu,\alpha} t_{\nu,\alpha}^\dagger t_{\nu,\alpha}^{\phantom{\dagger}}\quad ,
\end{equation}
where the sum runs over all sites $\nu$ of the effective lattice and the three triplet flavors $\alpha\in\{-1,0,+1\}$. The triplet operator $t_{\nu,\alpha}^{(\dagger)}$ destroys (creates) a triplet on site $\nu$ with flavor $\alpha$. The part $\mathcal{H}'/J^{\prime}$ can then be written as
\begin{equation}
 \frac{\mathcal{H}'}{J^{\prime}}=-\frac{3}{8} N + \sum_{\nu,\alpha} t_{\nu,\alpha}^\dagger t_{\nu,\alpha}^{\phantom{\dagger}} = -\frac{3}{8}N+ Q_{\rm Triplon} \; .
\end{equation}
The constant reflects the total energy of singlets on the $N/2$ isolated $b_{t^\prime}$-bonds.

The full spin Hamiltonian can now be recasted into
\begin{equation}
 \frac{\mathcal{H}}{J^{\prime}}=\tilde{E}_0+Q_{\rm Triplon} + \sum_{n=-3}^{3} T_n \;,
\end{equation}
where the $T_n$ operators create (destroy) $n$ triplets. Note that the operators $T_{\pm 3}$ arise from the four-spin interactions of the effective model, since these couple three dimers simultaneously. The pCUT maps this Hamiltonian, order by order in $x_\kappa$, to an effective Hamiltonian $\mathcal{H}_{\rm eff}$
 which commutes with $Q_{\rm Triplon}$, i.e.~the effective model is blockdiagonal in the number of triplons.
  
Here we focus on the effective one-triplon block in order to determine the one-triplon gap $\Delta/J^{\prime}$. To this end one calculates the one-triplon hopping amplitudes on appropriate clusters of the effective triangular lattice such that the results are correct in the thermodynamic limit.
The corresponding one-triplon hopping Hamiltonian is then diagonalized by Fourier transformation yielding the one-triplon dispersion $\omega(\vec{k})$ as well as the gap $\Delta\equiv\omega(\vec{k}=0)$ in units of $J^{\prime}$.
 
We performed the pCUT in two ways: First, we expand up to order 5 in all six expansion parameters $x_\kappa$ with  $\kappa\in\{1,2,3,\times,=,\parallel\}$. It it noteworthy that the specific gCUT values of the expansion parameters in terms of the ratios $t/U$ and $t^\prime/U$ can be inserted in the series after the pCUT has been performed. The pCUT can therefore be applied independently of the gCUT. Second, we reached order 6 by i) specifying the $J_\kappa$ explicitly with the corresponding gCUT values for fixed Hubbard parameters and ii) expressing all couplings $x_\kappa$ relative to a single expansion parameters $x$, e.g.~$x\equiv x_1$, $x_2=\phi_2 x_1$, $x_3=\phi_3 x_1$, and so on with $\phi_\kappa$ fixed by $t$, $t^\prime$, and $U$. The drawback is that the pCUT must be performed for each value of $t/U$ and $t^\prime/t$ individually.

In order to detect a possible second-order quantum phase transition, we use dlogPad\'e techniques to extrapolate the one-particle gap $\Delta$. Various extrapolants $\left[L,M\right]$ are constructed where $L$ denotes the order of the numerator and $M$ the order of the denominator. Explicitly, the dlogPad\'e extrapolation is based on the Pad\'{e} extrapolation of the logarithmic derivative of the one-triplon gap
\begin{equation}
 \left[\frac{d}{dx}\ln \Delta\right]_{[L,M]}:=\frac{P_{L}}{Q_M}\quad ,
 \label{eq:dlog}
\end{equation}
where $P_{L}$ and $Q_M$ are polynomials of order $L$ and $M$. Due to the derivative of the numerator in Eq.~\ref{eq:dlog} one requires $L+M=m-1$ where $m$ denotes the maximum perturbative order which has been calculated. The $\left[L,M\right]$ dlogPad\'e extrapolant is then given by
\begin{equation}
 \left[L,M\right] :=\exp\left(\int_0^x \frac{P_{L}(x')}{Q_M(x')} dx'\right)\quad .
 \label{eq:dlog2}
\end{equation}
In the case of a physical pole at $x_0$ one is able to determine the dominant power-law behaviour $|x-x_0|^{z\nu}$ close to $x_0$. The exponent $z\nu$ is then given by the residuum of $P_L/Q_M$ at $x=x_0$
\begin{equation}
 z\nu =\frac{P_{L}(x)}{\frac{d}{dx}Q_M(x)} |_{x=x_0}\quad .
 \label{eq:exp}
\end{equation}
For the problem under investigation we expect a quantum phase transition in the O(3) universality class having \mbox{$z=1$} for the dynamical critical exponent and \mbox{$\nu=0.7112(5)$} \cite{Campo02}. In general, one expects a better quality of the extrapolation with increasing perturbative order. 
%
\section{QMC}
\label{sec:qmc}
%
 At  half-band filling, the model of Eq. (\ref{Hubbard_model})  is amenable to sign problem free Quantum Monte Carlo  simulations.  
  Here we have adopted the projective auxiliary field QMC approach which is based on the identity:
  \begin{equation}
       \frac{\langle \Psi_0 | O  | \Psi_0 \rangle}{ \langle \Psi_0 | \Psi_0 \rangle}  = \lim_{\Theta \rightarrow \infty }  
       \frac{\langle \Psi_T |  {\rm e}^{-\Theta H /2 }  O {\rm e}^{-\Theta H /2 } | \Psi_T \rangle  }{  \langle \Psi_T |  {\rm e}^{-\Theta H }  | \Psi_T \rangle}.
  \end{equation}
  In the above, the trial wave function $  | \Psi_T \rangle $ corresponds to the ground state of the non-interacting problem and is assumed to be non-orthogonal to  the 
  ground state    $  | \Psi_0 \rangle $  of the interacting  Hamiltonian. Under this assumption,  propagation of  $ | \Psi_T \rangle $  along the imaginary time axis   will filter out the ground 
  state from the trial wave function.  There are many ways to implement the algorithm and the interested reader is referred to  Ref.~\onlinecite{Assaad08_rev} for a detailed review.   For the 
  present implementation, we  have opted for a  symmetric Trotter decomposition which minimizes the error due to the finite value of the time step adopted $\Delta  \tau$.   Typically  we have opted for 
  $\Delta \tau t = 0.1$ down to $  \Delta \tau t = 0.05$  for   simulations at large values of $U/t$. We have furthermore used a $\text{SU}(2)$-symmetric Hubbard Stratonovitch transformation. With this choice, the auxiliary field couples to the density such that $\text{SU}(2)$-spin symmetry is present  for each choice of the field.    Finally, let us comment on the value of $\Theta$ required to guarantee   convergence to the ground state.    We have carried out two types of simulations to at best determine the phase diagram.    On one side we have used the pinning field approach  \cite{Assaad13} to compute the magnetic moment.  As argued in 
Ref.~\onlinecite{Assaad13},  since this approach breaks the spin symmetry, very large values of the projection parameter are required to guarantee  convergence to the ground state.  The  pinning field simulations presented here are carried out at   $\Theta t = 320$.  For simulations where $\text{SU}(2)$-spin symmetry is present,  values of $\Theta t = 40$ suffice  for convergence.   Let us note that  the computational cost grows linearly with $\Theta $ such that reaching large projection parameters is not prohibitively  expensive.  
 
%
%
\section{Results}
\label{sec:results}
The  QMC  approach is at best suited to study the phase diagram  starting from weak coupling.  In contrast, the CUT excels in the strong coupling such that a combination of both methods has the potential of elucidating the nature of the phase diagram. Here we focus first on the values $t^\prime/t\in\{-1,-2\}$ giving representative results for the qualitatively different sequences of phases as a function of $U/t$. Afterwards, we study the breakdown of the VBS for continuously varying $t^\prime/t$.      

\subsection{SM to N\'eel: $t^\prime/t=-1$}

The case $t^\prime/t=-1$ corresponds to the $\pi$-flux square lattice. In the following we show by QMC that one finds a direct transition between the SM and a long-range ordered N\'eel state which is similar to the case $t^\prime=0$ where the model reduces to the Hubbard model on the isotropic honeycomb lattice. Compelling results \cite{Sorella12} on lattices up to $36 \times 36 $ unit cells as well as pinning field results of Ref.~\onlinecite{Assaad13} point to a direct transition  between the semi-metallic phase and antiferromagnetic insulator.  The phase transition  can be understood in terms of Gross Neveu  criticality. Simulations at $t^\prime = 0$ are facilitated by the C$_3$ symmetry   of the underlying triangular lattice, such that the Dirac cones  present at weak coupling are pinned to  interaction independent momenta.   Finite values of $t^\prime$ break  this symmetry and  the cones can meander.  For example in the absence of interactions,  the cones  will  meet  and annihilate at the $\Gamma$-point at $t^\prime/t=-3$ (see also Sect.~\ref{ssec:free}).  

Interestingly, the point $t^\prime/t = -1$ also possesses an enhanced symmetry. This C$_4$ symmetry becomes apparent when taking an adequate gauge choice setting all hopping matrix elements of the square lattice to $t\,{\rm e}^{{\rm i} \pi/4} $ and moving say clockwise around a plaquette. As a consequence, Dirac cones are pinned to wave vectors $(\pm \pi/2, \pm \pi/2)$.    Knowing where the nodes are located greatly simplifies the numerical calculations by QMC since it allows us to choose a set of lattice sizes where they are present. This set of lattice sizes  generically yield a smooth scaling to the thermodynamic limit.  
 
\begin{figure}
\begin{center}
\includegraphics*[width=\columnwidth]{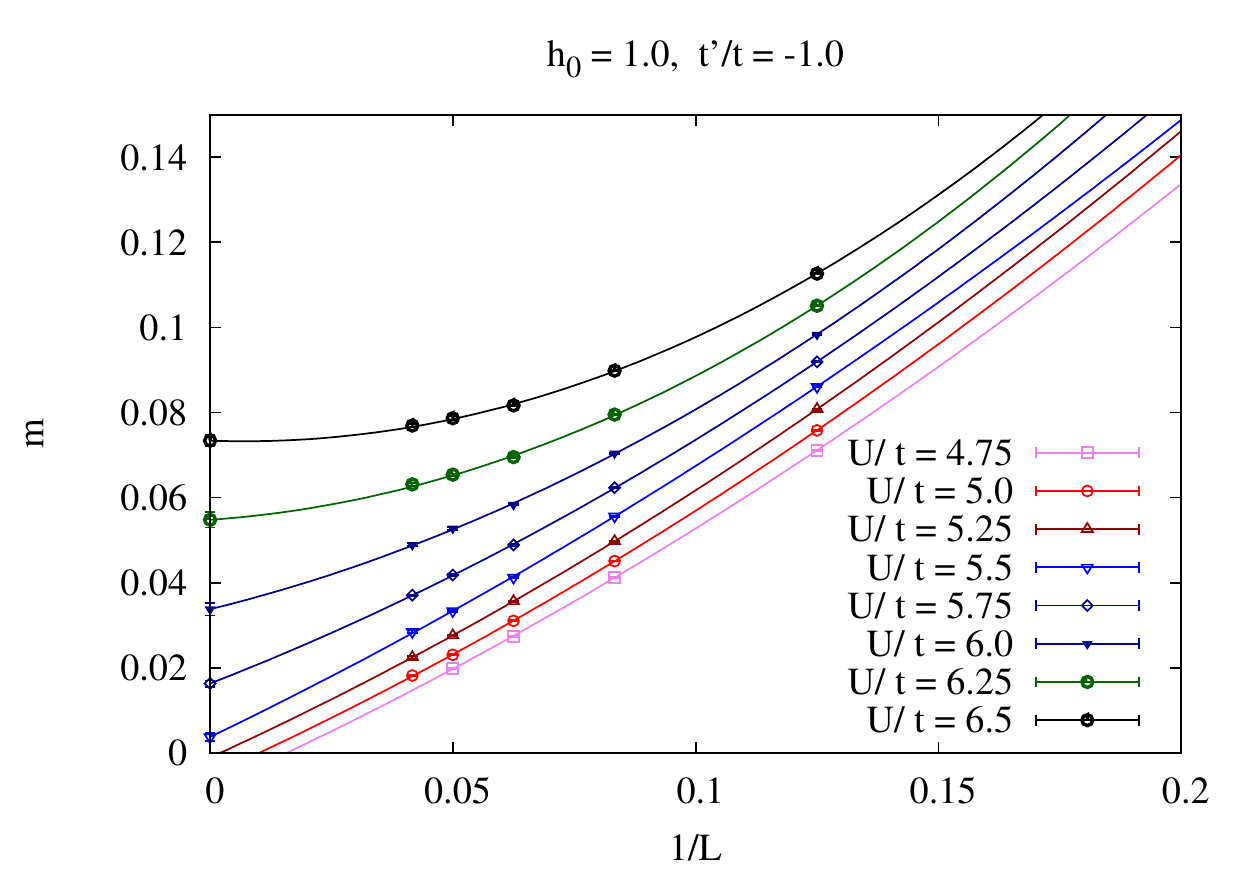} \\
\includegraphics*[width=\columnwidth]{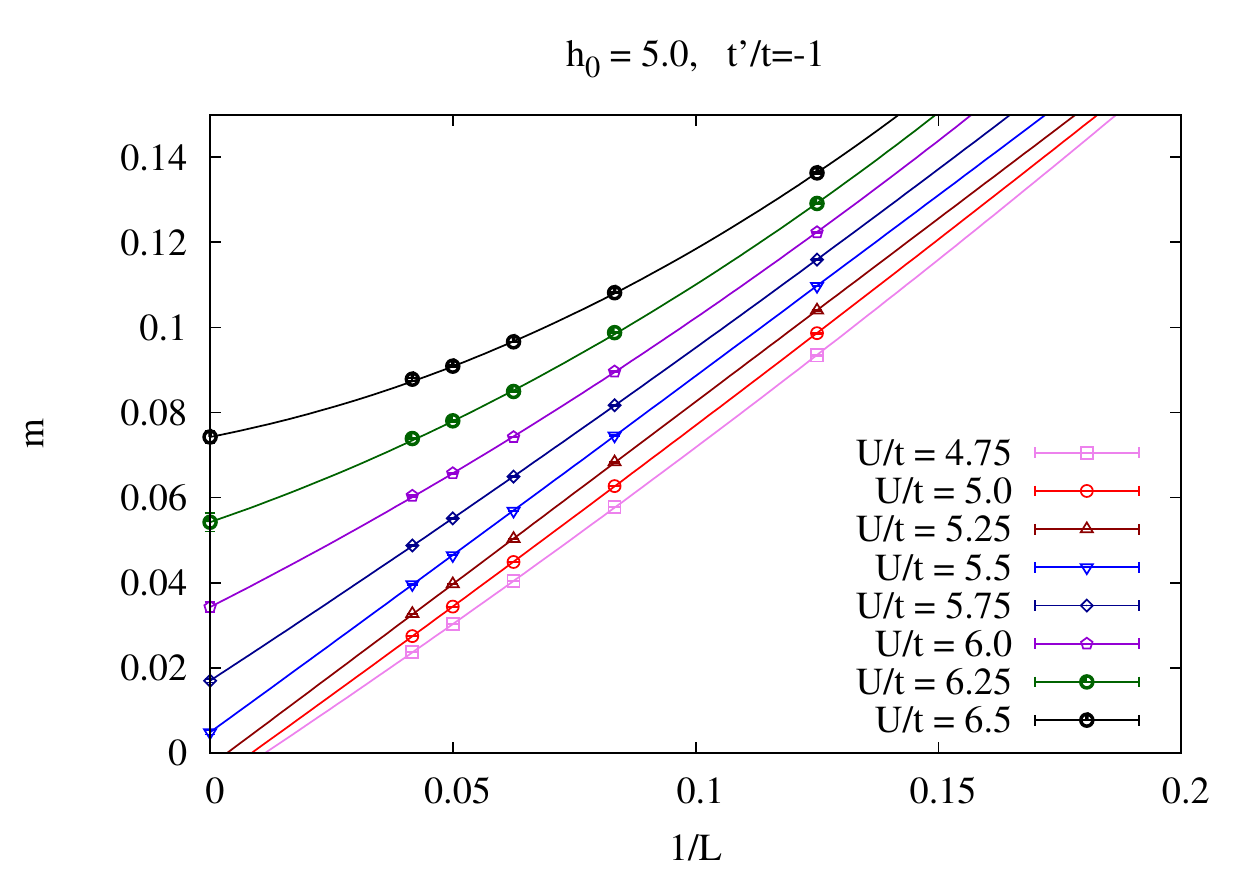} 
\end{center}
\caption{(Color online)  Pinning field QMC data for the induced magnetic moment $m$ as a function of $1/L$ at $t^\prime/t = -1$.  Here we have use a projection parameter $\Theta t = 320$. Both values of the pining field $h_0=1$ (top panel) and $h_0=5$ (bottom panel) lead to the same extrapolated value of the magnetization thus providing an internal check.} 
\label{fig:PiFlux_mag}
\end{figure}
\begin{figure}
\begin{center}
\includegraphics*[width=\columnwidth]{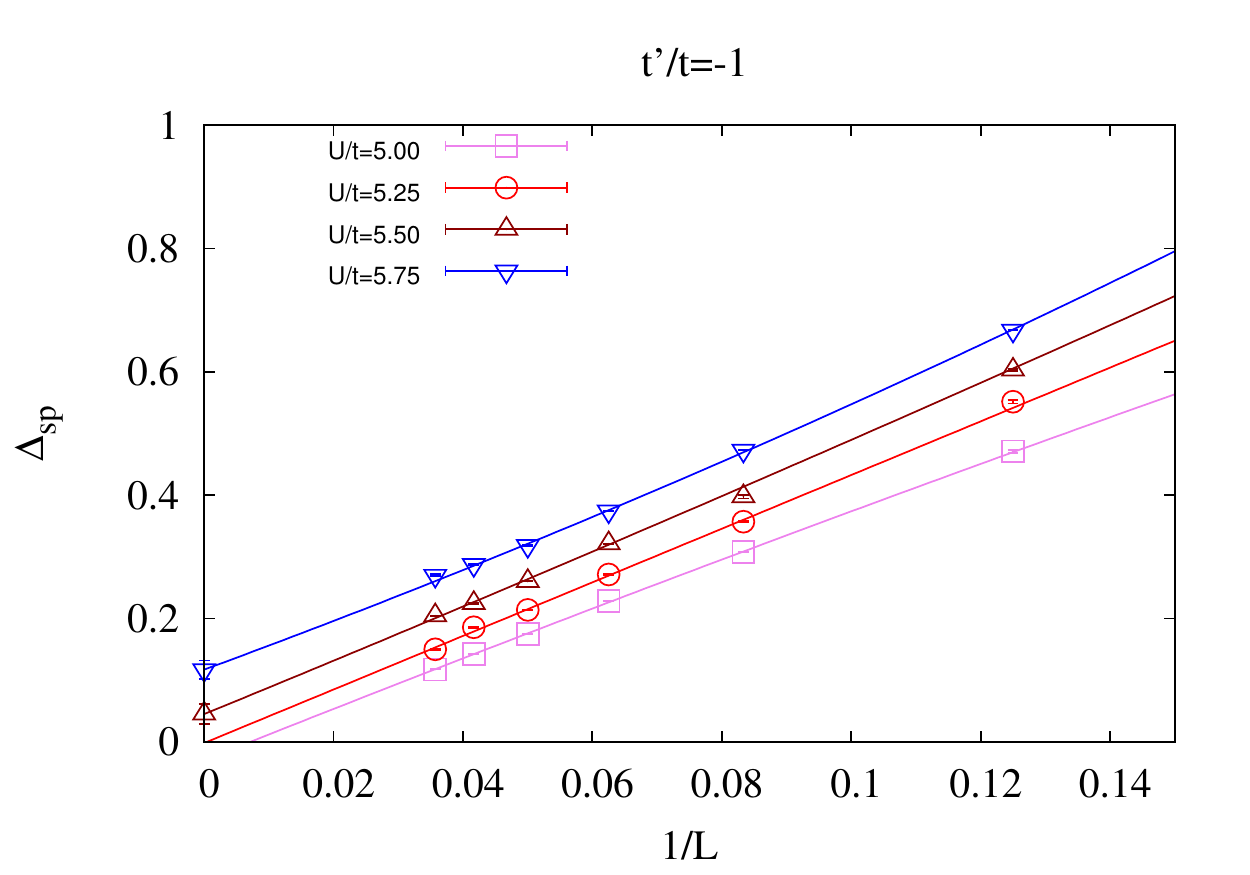}
\end{center}
\caption{(Color online) Single-particle gap $\Delta_{\rm sp}$ in units of $t$ as a function of $1/L$ for different ratios $U/t$ at $t^\prime/t=-1$ obtained by QMC. Lines are linear extrapolations up to the thermodynamic limit.} 
\label{fig:QP_tm1}
\end{figure}

Fig.~\ref{fig:PiFlux_mag} plots the magnetization as obtained from  the pinning field approach \cite{Assaad13}.  Here we add a local  magnetic field to the Hamiltonian  of Eq. (\ref{Hubbard_model})
\begin{equation}
\mathcal{H}_{\rm Local}  =    h_0  \left( n_{0,\uparrow} - n_{0,\downarrow}  \right)
\end{equation} 
and measure the induced  magnetic moment
\begin{equation}
       m    =  \frac{1}{2}  \frac{1}{N} \sum_{\vec{i}}  (-1)^{\vec{i}}     \langle n_{\vec{i},\uparrow} - n_{\vec{i},\downarrow} \rangle
\end{equation}
where $N $ corresponds to the total number of sites and $(-1)^{\vec{i}}$  takes the value $1$ ( $-1$)  on sublattice A (B). 
As apparent from Fig.~\ref{fig:PiFlux_mag}   both choices of the local field extrapolate to the {\it same}  value and point towards a  magnetic transition located in the interval 
$U_c/t \in \left] 5.25, 5.5 \right[$.   
To assess if the order triggers a mass gap, we have computed the single-particle gap, $\Delta_{\rm sp}$, at  the nodal point $\vec{K}$. As mentioned above, at $t^\prime/t = -1$  this nodal point is pinned  to an 
interaction independent value.    We extract  the single-particle gap form the imaginary time Green function  at the nodal point: 
\begin{equation}
	G(\vec{K},\tau)   = \sum_{\sigma,\alpha} \langle c^{\dagger}_{\vec{K}, \sigma,\alpha} (\tau) c_{\vec{K}, \sigma,\alpha} (0) \rangle 
\end{equation}
where $\alpha$ runs over the orbitals in the unit cell.  Asymptotically,  $G(\vec{K},\tau) \propto Z {\rm e}^{-\Delta_{\rm sp}  \tau} $ where $Z$ corresponds to the quasi-particle residue and 
$\Delta_{\rm sp} $ to the desired single-particle gap.  Fig.~\ref{fig:QP_tm1} plots this quantity  as a function of system size and as apparent   we reach the same conclusion as for the magnetization, namely that 
$U_c/t \in \left] 5.25, 5.5 \right[$.  Hence,   we will conclude that  the mass gap originates for the sublattice and time reversal symmetry breaking inherent to the magnetic ordering. 
Thus  at $t^\prime/t = -1$ our results are consistent with a direct transition from the semimetal to the magnetic insulator with N\'eel order.  

\begin{figure}
\begin{center}
\includegraphics*[width=\columnwidth]{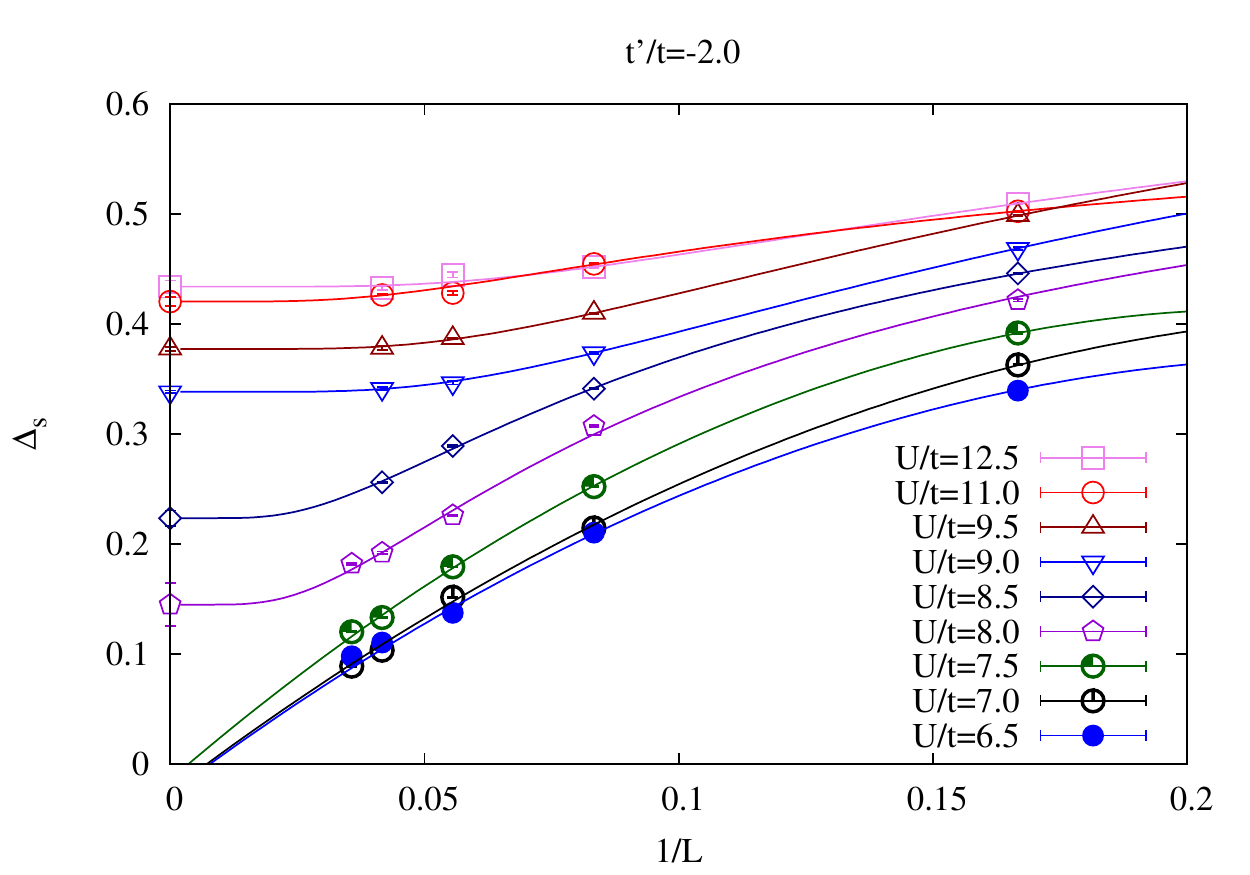}
\end{center}
\caption{(Color online)  Spin gap at $t^\prime/t = -2$ from QMC simulations.  The  extrapolation to the  thermodynamic limit is delicate. In the gapped  phase  we have  used the fitting form  $ a + b {\rm e}^{-\xi/L}$. This is  certainly an 
appropriate choice when the correlation length $\xi$ is smaller than the system size. If the data does not support this point of view, we have used a polynomial fit to extract the infinite volume value of the 
spin gap.} 
\label{fig:DS_QMC_tm2}
\end{figure}
\begin{figure}
\begin{center}
\includegraphics*[width=\columnwidth]{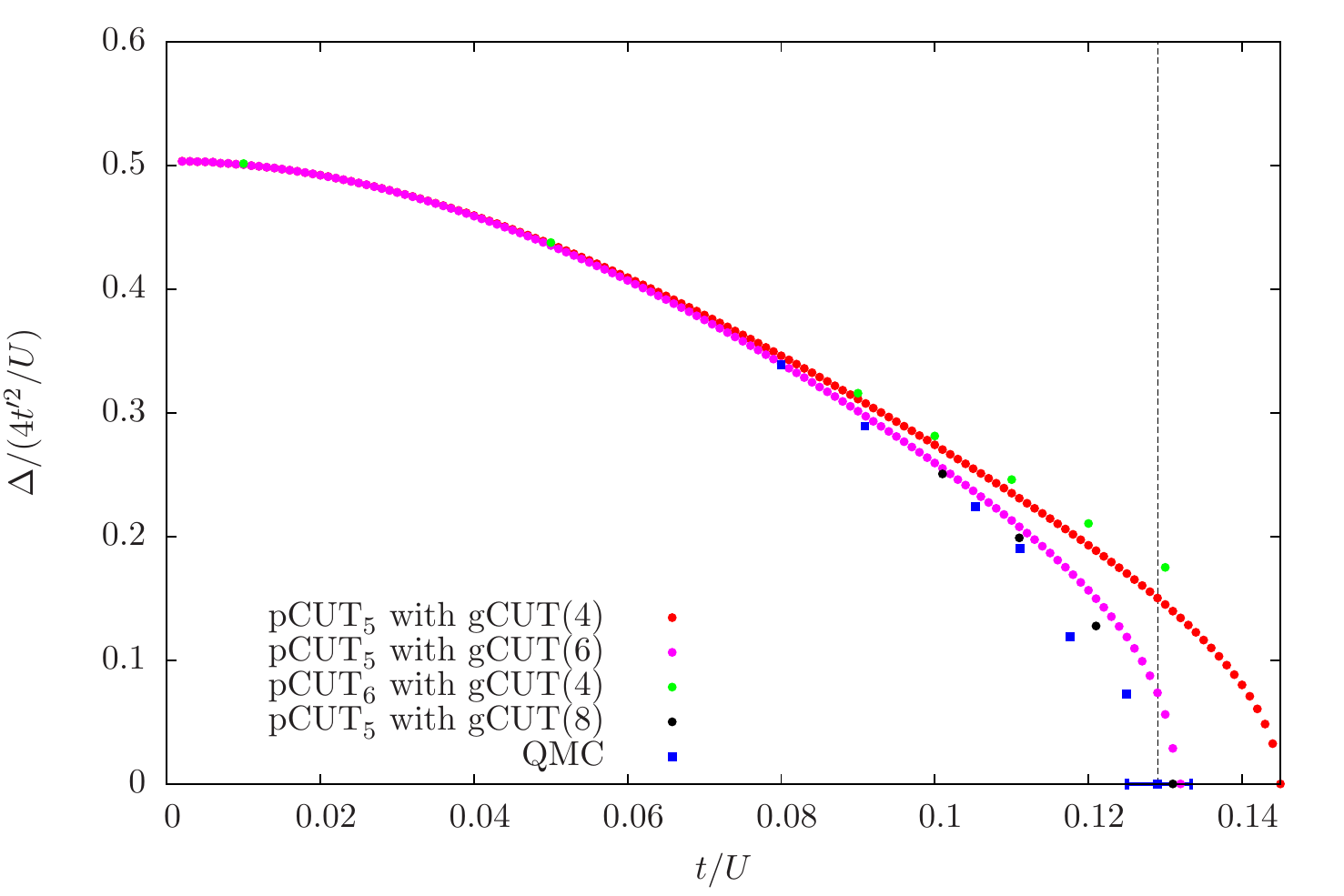}
\end{center}
\caption{(Color online) One-triplon gap $\Delta$ in units of $4t'^2/U$ as a function of $t/U$ at $t^\prime/t=-2$ obtained by CUTs (circles) and QMC (squares). For the CUT, we display different truncations of the gCUT($\nu$) with $\nu\in\{4,6,8\}$ and different maximal orders $n\in\{5,6\}$ of the pCUT series expansion. For the latter we have used dlogPad\'e extrapolants $[2,2]$ and $[2,3]$, respectively. The estimated quantum critical point is located at the vertical dashed line.} 
\label{fig:gapCompare_s-2}
\end{figure}
\begin{figure}
\begin{center}
\includegraphics*[width=\columnwidth]{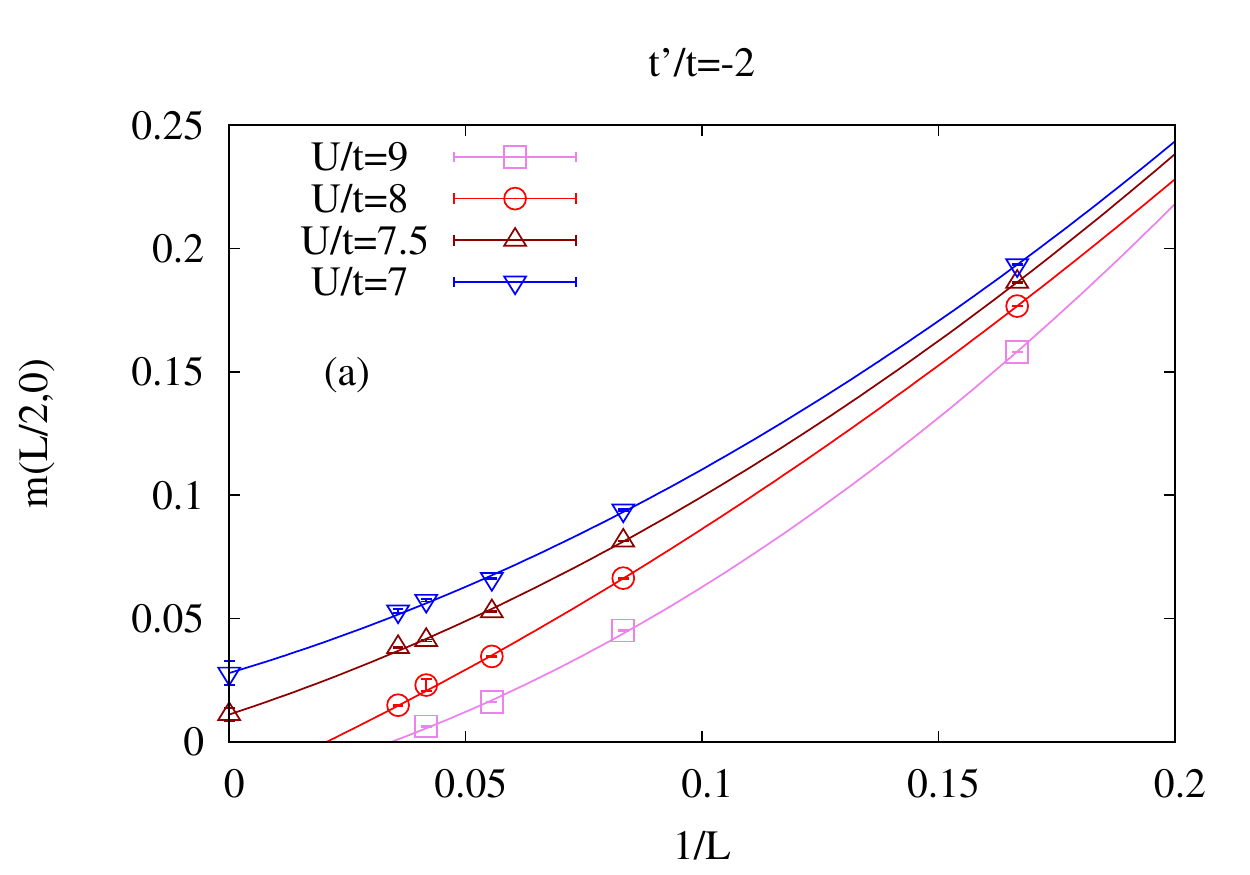}
\includegraphics*[width=\columnwidth]{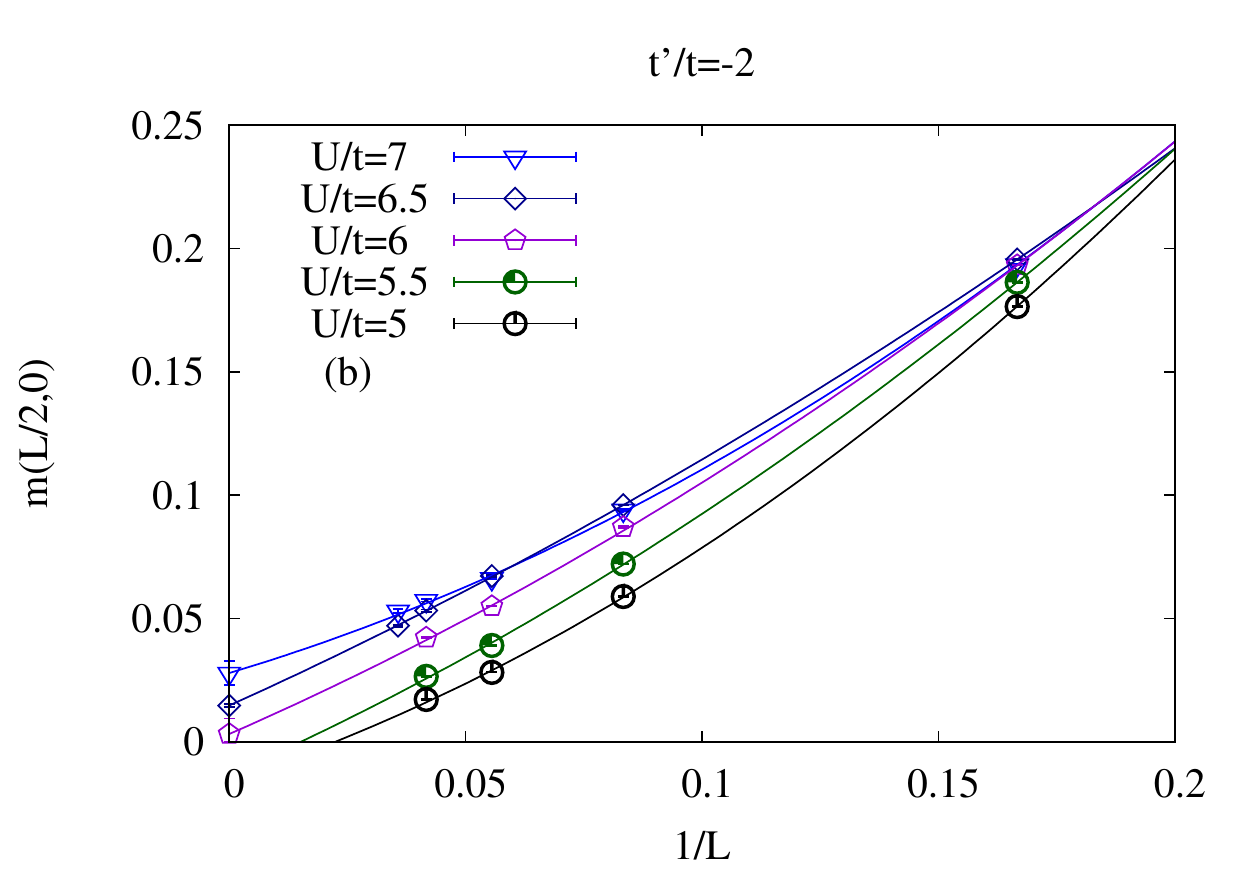}
\end{center}
\caption{(Color online) Real space spin-spin correlations at $t^\prime/t = -2$ from QMC simulations for (a) $U/t\geq 7$ and (b) $U/t\leq 7$. Here we  consider the largest distance along the x-axis on a $L\times L$ lattice.  We fit the data to the form \mbox{$a + b/L + c/L^2$}.} 
\label{fig:Spineq_X_tm2}
\end{figure}

\subsection{SM to N\'eel to VBS: $t^\prime/t=-2$}

Next we focus on $t^\prime/t=-2$. In contrast to the $\pi$-flux square lattice, the VBS is realized in the strong-coupling limit. We therefore deduce the phase diagram by QMC and CUTs yielding the presence of a N\'eel-ordered intermediate phase between SM and VBS. 

At $t^\prime/t = -2$, there is no symmetry which pins the Dirac points to specific values of the momenta such that a precise calculation of the single-particle gap with QMC is hard. In this  case it is more convenient to start from strong coupling. The VBS state is characterized by a finite spin gap to triplon excitations, which we can determine either with QMC in a very similar way as the single-particle gap or by the CUT approach yielding the one-triplon gap as a high-order series expansion in magnetic exchange couplings linked to the original Hubbard model by our gCUT approach. 

With QMC, we measure the imaginary time displaced spin-spin correlation function at the anti-ferrromagnetic wave vector and fit the tail of the QMC data to the form $Z_s {\rm e}^{-\Delta_{\rm s} \tau}$  where 
\begin{equation}
	\Delta_{\rm s} =  E_0(S=1) - E_0(S=0)
\end{equation}
is the energy difference between the \mbox{$S=1$} and \mbox{$S=0$} ground-state energies. Our QMC results for the spin gap are plotted in Fig.~\ref{fig:DS_QMC_tm2} and, as apparent, the VBS state survives down to $U_c/t \in \left] 7.5, 8.0 \right[ $.  Comparison with the corresponding CUT results are    
 displayed in Fig.~\ref{fig:gapCompare_s-2}. Most importantly, agreement is very good which also solidifies the interpretation of the QMC spin gap as originating from single-triplon excitations.

%
\begin{figure}
\begin{center}
\includegraphics*[width=\columnwidth]{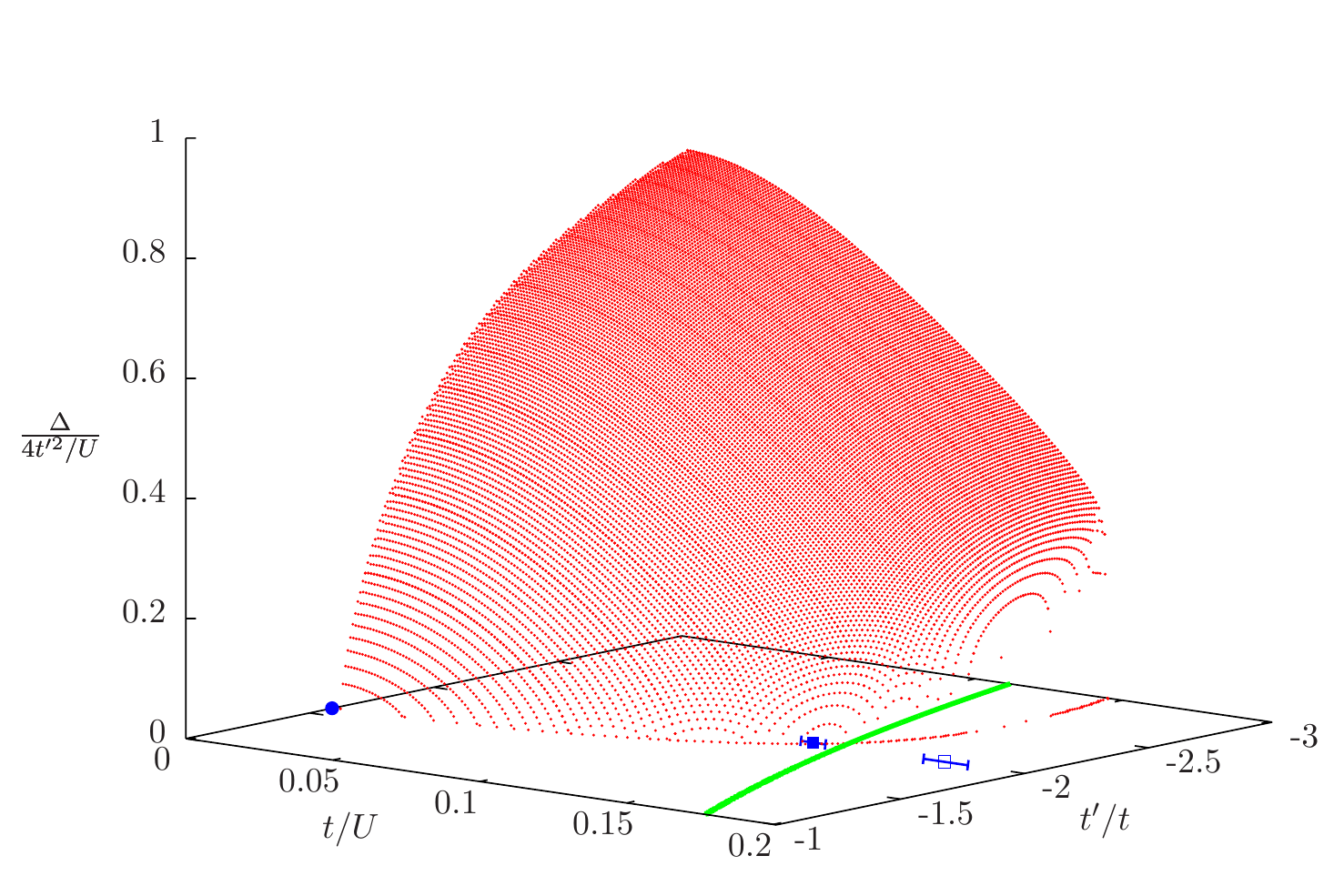}
\end{center}
\caption{(Color online) One-triplon gap $\Delta$ in units of $4t'^2/U$ as a function of $t^\prime/t$ and $t/U$ as obtained from CUT calculations (red points). Solid green line denotes the characteristic energy scale $t/W$. Symbols corresponds: i) Circle denotes quantum phase transition between N\'eel order and VBS in the strong-coupling limit $t/U\ll 1$ and ii) squares represent QMC results for $t^\prime/t=-1$ and $t^\prime/t=-2$ from this work.} 
\label{fig:phase_diagram}
\end{figure}

There  are two possible mean-field like scenarios which can account for the instability of the Dirac semi-metal to the VBS state.  Since there is no symmetry  reason to pin  the momenta of the Dirac points, they can meander as a function of the interaction strength, and meet at the $\Gamma$-point at $U_c$. This is precisely  what happens  in the absence of interactions at $t^\prime/t=-3$ (see also Sect.~\ref{ssec:free}).   Note that within this picture,  the density of states at the critical point is given by $\sqrt{\omega}$.  At the RPA level,  this does not lead to a weak coupling magnetic instability.  The second possibility is that there is an ordered phase separating the weak coupling Dirac semi-metal from the VBS. Our numerical QMC results as well as the fact that the VBS breaks down by triplon condensation support this point of view.  

We have computed the real space spin-spin correlations and extracted the local moment: 
\begin{equation}
	m(\vec{r})  = \sqrt{  \langle    \left( n_{\vec{r}+ \vec{r}_0,\uparrow} - n_{ \vec{r} + \vec{r}_0,\downarrow} \right)  \left(  n_{\vec{r}_0,\uparrow} - n_{\vec{r}_0,\downarrow} \right)  \rangle }. 
\end{equation}
Fig.~\ref{fig:Spineq_X_tm2}   plots the above quantity  ar $ \vec{r} = (L/2,0) $  on an $L \times L$ lattice. We choose the x-direction since the dimerization being along the y-axis we expect dominant spin-spin correlations 
along the former axis.  As apparent from Fig.~\ref{fig:Spineq_X_tm2} no order is detectable in the VBS phase for $U\geq 8 t$. Below the VBS phase, at $U =  7t $ for instant, the data supports    long range magnetic ordering which gives way to a paramagnetic phase below $ U = 6t$.   

Thus, at $t^\prime/t =-2$ the QMC data supports the following picture. First, the Dirac semi-metal  develops a mass gap due to the onset of anti-ferromagnetic correlations. This transition is expected to be in the Gross-Neveu universality class. At a slightly  larger value of $U/t$ this magnetically ordered phase gives way to the VBS. This transition does not involve fermonic degrees of freedom and is expected to belong to the O(3) universality class. 

The latter scenario is in full agreement with our CUT approach. First, the quantum phase transition between VBS and N\'eel order is located at $U\geq 8 t$, where we observe that our effective spin model nicely converges, i.e.~fermionic degrees of freedom can be well separated from magnetic degrees of freedom in this parameter regime. Second, the CUT approach detects the breakdown of the VBS at very similar values of $U/t$ by investigating the one-triplon gap with a momentum consistent with N\'eel order. Furthermore, also the critical exponent $z\nu$ of the one-triplon gap is consistent with the expected O(3) universality class as we elaborate on in the next section.

\subsection{Triplon condensation}

Next we study the breakdown of the VBS by triplon condensation for general values of $t^\prime/t$ by analysing the one-triplon gap deduced from the CUT approach. The corresponding values from extrapolating the one-triplon gap with dlogPad\'e extrapolation are shown in Fig.~\ref{fig:phase_diagram}. 

Let us start with the strong-coupling limit $t/U\ll 1$. Here the gCUT part is expected to be fully converged and the effective spin Hamiltonian essentially reduces to the $J$-$J^{\prime}$-model with $J/J^{\prime}=1/4$ (see Sect.~\ref{ssec:strong_coupling}). Nevertheless, this does not imply that analysing the series has to be simple, but in our case it works rather well although the obtained perturbative order is moderate. We find a critical point at $J/J^{\prime}\approx -1.6$ in good agreememt with QMC simulations of the (unfrustrated) $J$-$J^{\prime}$-model \cite{Wenzel08}. At the same time the critical exponent $z\nu$ is found to be $\approx 0.73$ which is only slightly larger than the expected one from the O(3) universality class being $\approx 0.7$. Note that this overshooting behaviour is rather typical for high-order series expansions.

The results for $t^\prime/t=-2$ from QMC and CUTs suggest that the breakdown of the VBS does always correspond to a softening of the one-triplon mode and a quantum phase transition inside the Mott insulator to a N\'eel-ordered state is expected. It is therefore interesting to investigate the associated critical line and check whether the critical exponent remains constant when varying $t^\prime/t$. 

Interestingly, this is not case. We observe that the critical exponent monotonically shrinks from $z\nu\approx 0.73$ at $t^\prime/t=-1.6$ down to almost zero when reducing to $t^\prime/t=-3$. Clearly, this unphysical behaviour should be attributed to uncertainties in the CUT approach, mostly from the first step using gCUT to separate spin and charge degrees of freedom. Indeed, for $t^\prime/t<-2$, the expected quantum phase transition between VBS and N\'eel order takes place at large values of $t/U>t/W$ where a separation of spin and charge degrees of freedom becomes challenging. We are therefore convinced that the current implementation of the gCUT breaks down for $t/U>t/W$ and one should not trust the results in this regime. Likely, the same effect is present already for $t^\prime/t\approx -2$ yielding still good estimates for the critical point but giving uncertainties in the more sensitive critical exponent. One can therefore track the one-triplon gap of the VBS quantitatively in a wide parameter regime inside the VBS phase, but quantum critical points are only well described for $-1.6<t^\prime/t<-2$.

%
%
\section{Conclusions}
\label{sec:conclusion}
%
%
We used QMC and CUTs to study the half-filled Hubbard model on a one-parameter family of vortex-full square lattices ranging from the isotropic case to weakly coupled Hubbard dimers. The ground-state phase diagram consists of a SM, a BI, a magnetically ordered N\'eel phase and a VBS. The breakdown of the SM is typically to an ordered N\'eel phase even in parameter regimes where a VBS phase is present at strong coupling. 

The CUT approach is done in a two-step process combining for the first time gCUTs and pCUTs. The gCUT is applied to separate spin and charge degrees of freedom non-perturbatively yielding an effective low-energy spin model for the Mott insulating phase. This becomes problematic if degrees of freedom start to overlap on the considered graphs which we indeed observe for the largest graphs and/or large values of $t^\prime/t$. It would be very fascinating (but challenging) to improve the generator of the CUT on the graphs in order to disentangle spin and charge states in parameter regimes where the lowest energy levels on graphs are still well described by spin states. In a second step, we derived high-order series expansions for the one-triplon gap inside the VBS phase by pCUTs. To this end the effective spin model is reduced to all couplings which fit on a single plaquette, i.e.~it contains two-spin and four-spin interactions. Interestingly, even in the parameter regime where the effective model seems to be well converged, our results indicate that a proper calculation of critical exponents is very challenging. This is most likely due to the fact that other magnetic couplings like six-spin interactions on double plaquettes (or others) become important in this $t/U$ regime.  

Physically, our results for the $\pi$-flux square lattice are in disagreement with Ref.~\onlinecite{Chang12}, since we do not find an intermediate spin-liquid phase but a direct transition between semi-metal and a N\'eel-ordered phase. Similar to the recent findings for the Hubbard model on the isotropic honeycomb lattice which result from numerical advancements \cite{Sorella12,Assaad13}, also our findings for the $\pi$-flux square lattices benefits from optimizations in the QMC simulations. Overall  our results  further support that Mott transitions of Dirac fermions are generically described by Gross-Neveu criticality where the mass gap is generated by symmetry  breaking fields.   

\section{Acknowledgements}
KPS acknowledges fruitful discussions with A. L\"auchli and H.-Y. Yang. FFA would like to thank L. Balents for discussions. We thank the LRZ-M\"unich and the J\"ulich Supercomputing center for generous allocation of CPU time. Financial support from the DFG grant AS120/9-1 is acknowledged.


\begin{thebibliography}{10}

\bibitem{Yan10} 
S. Yan, D.A. Huse, and S.R. White, Science {\bf 332}, 1173 (2011).

\bibitem{Depenbrock12} 
S. Depenbrock, I.P. McCulloch, and U. Schollwoeck, Phys. Rev. Lett. {\bf 109}, 067201 (2012).

\bibitem{Morita02} 
H. Morita, S. Watanabe, and M. Imada, J. Phys. Soc. Jpn. {\bf 71}, 2109 (2002).

\bibitem{Motrunich05}  
O.I. Motrunich, Phys. Rev. B {\bf 72}, 045105 (2005).

\bibitem{Kyung06} 
B. Kyung and A.M.S. Tremblay, Phys. Rev. Lett. {\bf 97}, 046402 (2006).

\bibitem{Sahebsara08} 
P. Sahebsara and D. S\'en\'echal, Phys. Rev. Lett. {\bf 100}, 136402 (2008).

\bibitem{Tocchio08} 
L.F. Tocchio, F. Becca, A. Parola, and S. Sorella, Phys. Rev. B {\bf 78}, 041101 (2008).

\bibitem{Yoshioka09} 
T. Yoshioka, A. Koga, and N. Kawakami, Phys. Rev. Lett. {\bf 103}, 036401 (2009).
\bibitem{Yang10} 
H.-Y. Yang, A.M. L\"auchli, F. Mila, and K.P. Schmidt, Phys. Rev. Lett. {\bf 105} 267204 (2010). 

\bibitem{Meng10} 
Z.Y. Meng, T.C. Lang, S. Wessel, F.F. Assaad, and A. Muramatsu, Nature {\bf 88}, 487 (2010).

\bibitem{Chang12}
C.-C. Chang and R.T. Scalettar, Phys. Rev. Lett. {\bf 109}, 026404 (2012)

\bibitem{Sorella12}
S. Sorella, Y. Otsuka, and S. Yunoki, Scientific Reports {\bf 2}, 992 (2012).

\bibitem{Assaad13}
F.F. Assaad and I.F. Herbut, Phys. Rev. X {\bf 3}, 031010 (2013).

\bibitem{Herbut09}
I.F. Herbut, V. Juri\ifmmode \check{c}\else \v{c}\fi{}i\ifmmode \acute{c}\else \'{c}\fi{}, and B. Roy, Phys. Rev. B {\bf 79}, 085116 (2009).

\bibitem{Ryu09}
S. Ryu, C. Mudry, and C.-Y. Hou, and C. Chamon, Phys. Rev. B {\bf 80}, 205319 (2009).

\bibitem{Wenzel08}
S. Wenzel, L. Bogacz, and W. Janke, Phys. Rev. Lett. {\bf 101}, 127202 (2008).

\bibitem{Assaad08_rev}
F.F. Assaad and H.G. Evertz, in Computational Many
Particle Physics, edited by H. Fehske, R. Schneider, and
A. Wei\ss e, eds., Lecture Notes in Physics {\bf 739}, 277, Springer Verlag, Berlin, (2008).

\bibitem{Yang11} 
H.-Y. Yang and K.P. Schmidt, Eur. Phys. Lett. {\bf 94}, 17004 (2001).

\bibitem{Yang12}
H.-Y. Yang, A.F. Albuquerque, S. Capponi, A. Laeuchli, and K.P. Schmidt, New J. Phys. {\bf 14}, 115027 (2012).
\bibitem{Dusuel10}
S. Dusuel, M. Kamfor, K.P. Schmidt, R. Thomale, and J. Vidal, Phys. Rev. B {\bf 81}, 064412 (2010).

\bibitem{Wegner94} 
F. Wegner, Ann. Phys. (Leipzig) {\bf 77}, 3 (1994).

\bibitem{Glazek93} 
S.D. G{\l}azek and K.G. Wilson, Phys. Rev. D {\bf 48}, 5863 (1993).

\bibitem{Glazek94} 
S.D. G{\l}azek and K.G. Wilson, Phys. Rev. D {\bf 49}, 4214 (1994).

\bibitem{Knetter00} 
C. Knetter C and G.S. Uhrig, Eur. Phys. J. B {\bf 13}, 209 (2000)

\bibitem{Mielke98} 
A. Mielke, Eur. Phys. J. B. {\bf 5}, 605 (1998).

\bibitem{Knetter03} 
C. Knetter, K.P. Schmidt, and G.S. Uhrig, J. Phys. A {\bf 36}, 7889 (2003).

\bibitem{MacDonald88} 
A.-H. MacDonald, S.M. Girvin, and D. Yoshioka, Phys. Rev. B {\bf 37}, 9753 (1988).	

\bibitem{Reischl04} 
A. Reischl, E. M\"uller-Hartmann, and G.S. Uhrig, Phys. Rev. B {\bf 70} 245124 (2004).

\bibitem{Hamerla10} 
S.A. Hamerla, S. Duffe, and G.S. Uhrig, Phys. Rev. B {\bf 82}, 235117 (2010).

\bibitem{Schmidt03}
K.P. Schmidt and G.S. Uhrig, Phys. Rev. Lett. {\bf 90}, 227204 (2003).

\bibitem{Campo02}
M. Campostrini, M. Hasenbusch, A. Pelissetto, P. Rossi, and E. Vicari, Phys. Rev. B {\bf 65}, 144520 (2002).

\bibitem{Schnalle2010} 
R. Schnalle and J. Schnack, Int. Rev. Phys. Chem. {\bf 29}, 403 (2010).

\end{thebibliography}
\end{document}